%% file: ms.tex
\documentclass[useAMS,usenatbib,usegraphicx]{mn2e}
\usepackage{graphicx}
\usepackage[breaklinks,colorlinks,citecolor=black,linkcolor=black,urlcolor=black]{hyperref}

\newcommand*{\mysub}[2]{\ensuremath{#1_{\mathrm{#2}}}}
\newcommand*{\unit}[1]{\ensuremath{\mathrm{\, #1}}}
\newcommand*{\sw}[1]{{\sc #1}}
\newcommand*{\phmin}{\hspace{1.9ex}} 

\newcommand*{\Omegam}{\mysub{\Omega}{m}}
\newcommand*{\Omegab}{\mysub{\Omega}{b}}
\newcommand*{\Omegal}{\mysub{\Omega}{\Lambda}}
\newcommand*{\Omegade}{\mysub{\Omega}{DE}}
\newcommand*{\fbar}{\mysub{f}{b}}
\newcommand*{\fgas}{\mysub{f}{gas}}
\newcommand*{\sigmaf}{\ensuremath{\sigma_{\hspace{-0.25ex}f}}}

\newcommand*{\Mgas}{\mysub{M}{gas}}

\newcommand*{\LCDM}{\ensuremath{\Lambda}CDM}
\newcommand*{\rhocr}{\mysub{\rho}{cr}}
\newcommand*{\mproton}{\mysub{m}{p}}
\newcommand*{\mH}{\mysub{m}{H}}
\newcommand*{\mHe}{\mysub{m}{He}}
\newcommand*{\nelec}{\mysub{n}{e}}
\newcommand*{\nH}{\mysub{n}{H}}
\newcommand*{\NH}{\mysub{N}{H}}
\newcommand*{\nHe}{\mysub{n}{He}}
\newcommand*{\ntot}{\mysub{n}{tot}}
\newcommand*{\YHe}{\mysub{Y}{He}}

\newcommand*{\dA}{\mysub{d}{A}}
\newcommand*{\dL}{\mysub{d}{L}}
\newcommand*{\atr}{\mysub{a}{tr}}
\newcommand*{\wet}{\mysub{w}{et}}

\newcommand*{\second}{\unit{s}}

\newcommand*{\Ms}{\unit{Ms}}

\newcommand*{\cm}{\unit{cm}}
\newcommand*{\km}{\unit{km}}
\newcommand*{\kpc}{\unit{kpc}}
\newcommand*{\Mpc}{\unit{Mpc}}

\newcommand*{\keV}{\unit{keV}}
\newcommand*{\Msun}{\ensuremath{\, M_{\odot}}}

\newcommand*{\E}[1]{\ensuremath{\times 10^{#1}}}

\newcommand*{\ltsim}{\ {\raise-.75ex\hbox{$\buildrel<\over\sim$}}\ }
\newcommand*{\gtsim}{\ {\raise-.75ex\hbox{$\buildrel>\over\sim$}}\ }
\newcommand*{\proptosim}{\ {\raise-.75ex\hbox{$\buildrel\propto\over\sim$}}\ }
\newcommand*{\dnorm}{\ensuremath{\mathcal{N}}}
\newcommand*{\dunif}{\ensuremath{\mathcal{U}}}

\newcommand*{\secref}{Section}
\newcommand*{\appref}{Appendix}
\newcommand*{\eqnref}{Equation}
\newcommand*{\figref}{Figure}
\newcommand*{\tabref}{Table}

\newcommand*{\Chandra}{{\it Chandra}}
\newcommand*{\Planck}{{\it Planck}}

\newcommand*{\xsmod}[1]{\ensuremath{\widetilde{#1}}}
\newcommand*{\xsrs}{\mysub{\xsmod{r}}{s}}

\defcitealias{Allen0706.0033}{A08}
\newcommand*{\arsemf}{\citetalias{Allen0706.0033}}
\defcitealias{Navarro9611107}{NFW}
\newcommand*{\NFW}{\citetalias{Navarro9611107}}
\defcitealias{von-der-Linden1208.0597}{WtG}
\newcommand*{\wtg}{\citetalias{von-der-Linden1208.0597}}

\newcommand*{\morphpaper}{Paper~I}

\newcommand*{\profilespaper}{Paper~III}

\begin{document}

\title[Relaxed Galaxy Clusters: Cosmological Constraints]{Cosmology and Astrophysics from Relaxed Galaxy Clusters II: Cosmological Constraints}

\author[A. B. Mantz et al.]{A. B. Mantz,$^{1,2}$\thanks{E-mail: \href{mailto:amantz@kicp.uchicago.edu}{\tt amantz@kicp.uchicago.edu}} {} 
  S. W. Allen,$^{3,4,5}$
  R. G. Morris,$^{3,5}$
  D. A. Rapetti,$^6$ \newauthor
  D. E. Applegate,$^7$
  P. L. Kelly,$^8$
  A. von der Linden,$^{3,4,6}$
  R. W. Schmidt$^9$\\
  $^1$Department of Astronomy and Astrophysics, University of Chicago, 5640 South Ellis Avenue, Chicago, IL 60637, USA\\
  $^2$Kavli Institute for Cosmological Physics, University of Chicago, 5640 South Ellis Avenue, Chicago, IL 60637, USA\\
  $^3$Kavli Institute for Particle Astrophysics and Cosmology, Stanford University, 452 Lomita Mall, Stanford, CA 94305, USA\\
  $^4$Department of Physics, Stanford University, 382 Via Pueblo Mall, Stanford, CA 94305, USA\\
  $^5$SLAC National Accelerator Laboratory, 2575 Sand Hill Road, Menlo Park 94025, CA, USA\\
  $^6$Dark Cosmology Centre, Niels Bohr Institute, University of Copenhagen, Juliane Maries Vej 30, 2100 Copenhagen, Denmark\\
  $^7$Argelander-Institute for Astronomy, Auf dem H\"ugel 71, D-53121 Bonn, Germany\\
  $^8$Department of Astronomy, University of California, Berkeley, CA 94720, USA\\
  $^9$Astronomisches Rechen-Institut, Zentrum f\"ur Astronomie der Universit\"at Heidelberg, M\"onchhofstrasse 12-14, 69120 Heidelberg, Germany
}
\date{Accepted 2014 February 24. Received 2014 February 08; in original form 2013 December 20}

\pagerange{\pageref{firstpage}--\pageref{lastpage}} \pubyear{2014}
\maketitle
\label{firstpage}

\begin{abstract}
  This is the second in a series of papers studying the astrophysics and cosmology of massive, dynamically relaxed galaxy clusters. The data set employed here consists of \Chandra{} observations of 40 such clusters, identified in a comprehensive search of the \Chandra{} archive for hot ($kT\gtsim 5\keV$), massive, morphologically relaxed systems, as well as high-quality weak gravitational lensing data for a subset of these clusters. Here we present cosmological constraints from measurements of the gas mass fraction, \fgas{}, for this cluster sample. By incorporating a robust gravitational lensing calibration of the X-ray mass estimates, and restricting our measurements to the most self-similar and accurately measured regions of clusters, we significantly reduce systematic uncertainties compared to previous work. Our data for the first time constrain the intrinsic scatter in \fgas{}, $7.4\pm2.3$ per cent in a spherical shell at radii 0.8--1.2\,$r_{2500}$ ($\sim1/4$ of the virial radius), consistent with the expected level of variation in gas depletion and non-thermal pressure for relaxed clusters. From the lowest-redshift data in our sample, five clusters at $z<0.16$, we obtain a constraint on a combination of the Hubble parameter and cosmic baryon fraction, $h^{3/2}\,\Omegab/\Omegam = 0.089\pm0.012$, that is insensitive to the nature of dark energy. Combining this with standard priors on $h$ and $\Omegab h^2$ provides a tight constraint on the cosmic matter density, $\Omegam=0.27\pm0.04$, which is similarly insensitive to dark energy. Using the entire cluster sample, extending to $z>1$, we obtain consistent results for \Omegam{} and interesting constraints on dark energy: $\Omegal=0.65^{+0.17}_{-0.22}$ for non-flat \LCDM{} (cosmological constant) models, and $w=-0.98\pm0.26$ for flat models with a constant dark energy equation of state. Our results are both competitive and consistent with those from recent cosmic microwave background, type Ia supernova and baryon acoustic oscillation data. We present constraints on more complex models of evolving dark energy from the combination of \fgas{} data with these external data sets, and comment on the possibilities for improved \fgas{} constraints using current and next-generation X-ray observatories and lensing data.
\end{abstract}


\clearpage
\section{Introduction} \label{sec:cosintro}

The matter budget of massive clusters of galaxies, and specifically the ratio of gas mass to total mass, provides a powerful probe of cosmology (\citealt{White1993Natur.366..429W, Sasaki9611033, Pen9610090, Allen0205007, Allen0405340, Allen0706.0033}; \citealt*{Allen1103.4829}, and references therein). In these systems, the mass of hot, X-ray emitting gas far exceeds that in colder gas and stars (e.g.\ \citealt{Lin0408557}; \citealt*{Gonzalez0705.1726}; \citealt{Giodini0904.0448, Dai0911.2230, Leauthaud1104.0928, Behroozi1207.6105}), and the gas mass fraction, \fgas{}, is expected to approximately match the cosmic baryon fraction, $\Omegab/\Omegam$ (\citealt{Borgani0906.4370}, and references therein). Hydrodynamic simulations of cluster formation indicate that the gas mass fraction at intermediate to large cluster radii should have a small cluster-to-cluster scatter and evolve little or not at all with redshift (\citealt{Eke9708070, Kay0407058, Crain0610602}; \citealt*{Nagai0609247}; \citealt{Young1007.0887, Battaglia1209.4082, Planelles1209.5058}). Increasingly, as simulations have incorporated more accurate models of baryonic physics in clusters, in particular modeling the effects of feedback from active galactic nuclei (AGN) in cluster cores (e.g.\ \citealt{McNamara0709.2152}), they have become able to more reliably predict the baryonic depletion of clusters relative to the Universe as a whole. Combining such predictions with measurements of cluster \fgas{} and constraints on $\Omegab$, for example from cosmic microwave background (CMB) or Big Bang Nucleosynthesis (BBN) data and direct estimates of the Hubble parameter, provides a uniquely simple and robust method to constrain the cosmic matter density, $\Omegam$. The pioneering work of \citet{White1993Natur.366..429W} was among the first to show a clear preference for a low-density universe with $\Omegam\sim0.3$, a result which cluster \fgas{} data continue to support with ever greater precision (\citealt{Allen0205007, Allen0405340, Allen0706.0033, Ettori0211335, Ettori0904.2740}; \citealt*{Rapetti0409574}), and which has been corroborated by a variety of independent cosmological data (e.g.\ \citealt{Percival0608635, Percival0907.1660, Spergel0603449,  Kowalski0804.4142, Mantz0709.4294, Mantz0909.3098, Dunkley0803.0586, Vikhlinin0812.2720, Rozo0902.3702, Blake1108.2635, Komatsu1001.4538, Hinshaw1212.5226, Suzuki1105.3470, Anderson1303.4666, Benson1112.5435, Hasselfield1301.0816, Planck1303.5076}).

Given a bound on the evolution of \fgas{} from theory or simulations, the apparent evolution in \fgas{} values measured from X-ray data can also provide important constraints on the cosmic expansion history and dark energy \citep{Sasaki9611033, Pen9610090}. This sensitivity follows from the fact that derived \fgas{} values depend on a combination of luminosity and angular diameter distances to the observed clusters, analogously to the way that type Ia supernova probes of cosmology exploit the distance dependence of the luminosity inferred from an observed flux. \citet{Allen0405340} provided the first detection  of the acceleration of the cosmic expansion from \fgas{} data, and more recently expanded and improved their analysis (\citealt{ Allen0706.0033}, hereafter \arsemf{}; see also \citealt{LaRoque0604039, Ettori0904.2740}).

A key requirement for this work is that systematic biases and unnecessary scatter in the \fgas{} measurements be avoided. This can be achieved by limiting the analysis to the most massive, dynamically relaxed clusters available. The restriction to relaxed systems minimizes systematic biases due to departures from hydrostatic equilibrium and substructure, as well as scatter due to these effects, asphericity, and projection \citep{Rasia0602434, Nagai0703661, Battaglia1209.4082}. Similarly, using the most massive clusters minimizes residual systematic uncertainties associated with details of the hydrodynamic simulations, and simplifies the analysis by restricting it to those clusters for which \fgas{} is expected to have the smallest variation with mass or redshift, and the smallest intrinsic scatter \citep{Eke9708070, Kay0407058, Crain0610602, Nagai0609247,  Stanek0910.1599, Young1007.0887, Borgani0906.4370, Battaglia1209.4082, Planelles1209.5058, Sembolini1207.4438}. Moreover, the most massive clusters at a given redshift will be the brightest at X-ray wavelengths and require the shortest observing times.

This paper is the second of a series in which we study the astrophysics and cosmology of the most massive, relaxed galaxy clusters. The first installment (Mantz et~al.\ 2014, in preparation, hereafter \morphpaper{}) presents a procedure for identifying relaxed clusters from X-ray data based on their morphological characteristics, and identifies a suitable sample from a comprehensive search of archival \Chandra{} data. In future work (\profilespaper), we will investigate the astrophysical implications of our analysis of these clusters. This paper presents the cosmological constraints that follow from measurements of \fgas{} for the cluster sample.

Our work builds directly on that of \citet{Allen0205007, Allen0405340, Allen0706.0033}. Among our methodological improvements, three stand out as particularly important. First, the selection of target clusters has been automated (\morphpaper{}), enabling straightforward application to large samples. Second, the cosmological analysis uses gas mass fractions measured in spherical shells at radii near $r_{2500}$,\footnote{Defined as the radius within which the mean cluster density is 2500 times the critical density of the Universe at the cluster's redshift.} rather than \fgas{} integrated at all radii $<r_{2500}$. The exclusion of cluster centers from this measurement significantly reduces the corresponding theoretical uncertainty in  gas depletion from hydrodynamic simulations.\footnote{Improvements in the simulated physics, particularly the inclusion of feedback processes, have also been important in reducing this uncertainty (e.g.\ \citealt{Battaglia1209.4082, Planelles1209.5058}).} Third, the availability of robust mass estimates for the target clusters from weak gravitational lensing \citep[][hereafter collectively Weighing the Giants, or \wtg{}]{von-der-Linden1208.0597, Kelly1208.0602, Applegate1208.0605} allows us to directly calibrate any bias in the mass measurements from X-ray data, for example due to departures from hydrostatic equilibrium (e.g.\ \citealt{Rasia0602434, Nagai0609247, Battaglia1209.4082}) or instrument calibration (Applegate et~al., in preparation). In addition, our procedure employs blind analysis techniques (deliberate safeguards against observer bias) including hiding measured gas mass and total mass values until all analysis of the individual clusters was complete.

\secref~\ref{sec:cosmodata} reviews the selection of our cluster sample and basic X-ray data reduction (more fully described in \morphpaper{}), as well as the additional analysis steps required to derive \fgas{}. The resulting \fgas{} measurements are presented in \secref~\ref{sec:fgresults}. The cosmology and cluster models we fit to the data are described in \secref~\ref{sec:model}, and \secref{}~\ref{sec:cosmores} presents the cosmological results. \secref~\ref{sec:changes} summarizes the differences between our work and \arsemf{} (also discussed throughout \secref{}s~\ref{sec:cosmodata} and \ref{sec:model}) and compares their cosmological constraints. In \secref~\ref{sec:future}, we discuss the potential for further improvements in \fgas{} constraints from future observing programs targeting clusters discovered in upcoming surveys. We conclude in \secref~\ref{sec:cosmosummary}.

For the cosmology-dependent quantities presented in figures and tables, we adopt a reference flat \LCDM{} model with Hubble parameter $h=H_0/100\km\second^{-1}\Mpc^{-1}=0.7$, matter density with respect to the critical density $\Omegam=0.3$, and dark energy (cosmological constant) density $\Omegal=0.7$. However, our cosmological constraints are independent of the particular choice of reference (\arsemf{} and \secref~\ref{sec:model}).

\section{X-ray Data and Analysis} \label{sec:cosmodata}

\subsection{Cluster Sample} \label{sec:sample}

The data set employed here is limited to the most dynamically relaxed, massive clusters known. This restriction is critical for minimizing systematic scatter in the degree of non-thermal pressure in clusters, scatter due to global asymmetry and projection effects, and theoretical uncertainty in the implementation of relevant hydrodynamical simulations, any of which would weaken the final cosmological constraints.

Our selection of massive, relaxed clusters is described in detail in \morphpaper{}, and we provide only a short summary here. In \morphpaper{}, we introduce a set of morphological quantities which can be measured automatically from X-ray imaging data, as well as criteria for identifying relaxed clusters based on these measurements. In brief, the morphological test is based on (1) the sharpness of the peak in a cluster's surface brightness profile, (2) the summed distances between centers of neighboring isophotes (similar in spirit to centroid variance), and (3) the average distance between the centers of these isophotes and a global measure of the cluster center (a measure of global asymmetry). The isophotes referred to in (2) and (3) typically cover the radii $0.25<r/r_{2500}<0.8$, a range where the signal to noise ratio is generally adequate, but which deliberately excludes the innermost regions, where complex structure (e.g.\ associated with sloshing or AGN-induced cavities) is ubiquitous, even in the most relaxed clusters \citep{McNamara0709.2152, Markevitch0701821}.

This algorithm was run over a large sample of clusters ($>300$) for which archival \Chandra{} data were available as of February 1, 2013 to generate an initial candidate list. Two additional cuts were then applied. First, to identify the most massive systems, clusters for which the global temperature $kT<4\keV$ (either as measured previously in the literature or estimated from X-ray luminosity--temperature scaling relations, e.g.\ \citealt{Mantz0909.3099}) were eliminated. Note that our final temperature requirement, $kT \geq 5\keV$ in the relatively isothermal part of the temperature profile, was enforced later, using our own measurements (specifically, the projected, global temperature measured in \secref~\ref{sec:projct}) and the most recent \Chandra{} calibration information. Second, we identified for each cluster a central region vulnerable to the aforementioned morphological complexities, which was excluded from the mass measurement procedure (\secref~\ref{sec:projct}). This circular region has a minimum radius of 50\kpc{} (in our reference cosmology), but can be larger if there are visible disturbances in the cluster gas (e.g.\ clear cold fronts, which the morphology algorithm may not recognize if they are sufficiently symmetric in appearance or if they occupy sufficiently small cluster radii; see \tabref~\ref{tab:targets}).\footnote{Using a fixed metric radius for the minimum exclusion region is arguably unnecessarily conservative at high redshift, given that the region will extend out to much lower densities relative to the critical density. For the two $z>1$ clusters in our sample, 3C186 and CL\,J1415.2+3612, we have therefore reduced the minimum exclusion radius to 25\kpc. In addition, there are a small number of cases where we excluded particular position angle ranges at all radii from our analysis, as in \arsemf{}. These are listed in \tabref~\ref{tab:targets}.} Clusters for which this excluded region encompassed the brightest isophote identified in the morphology analysis (i.e.\ radii $\gtsim 0.25\, r_{2500}$) were removed from the sample.

Beyond the considerations described above and in \morphpaper{}, we eliminated three additional clusters from the final sample:
\begin{enumerate}
\item Abell 383: Our surface brightness profile for this cluster (\secref~\ref{sec:projct}) displays an unusual flattening between $\sim225$ and 400 arcsec, before again decreasing at large radii. We can identify no discrete sources in the X-ray data responsible for this. There is a concentration of red galaxies at approximately these radii northeast of the cluster \citep{Zitrin1108.4929}. However, an azimuthally resolved analysis of the X-ray surface brightness (in $60^\circ$ sectors) appears to show the excess extending over $\sim3/4$ of azimuths, albeit at lower significance. Lacking a good explanation for the source of this apparent excess emission, we have removed the cluster from our sample.
\item MACS\,J0326.8$-$0043: This cluster satisfies all of our criteria for selection, but the existing data are too shallow to constrain the temperature profile at $r_{2500}$, a requirement for our \fgas{} measurement.
\item MACS\,J1311.0$-$0311: The spectral background in the available data does not appear to be well described by the associated \Chandra{} blank-sky field. In particular, an excess of hard emission persists after background subtraction. Rather than attempting to model and subtract this excess, we have opted to remove the cluster from our sample.
\end{enumerate}

The final sample of 40 hot, relaxed clusters used in this work appears in \tabref~\ref{tab:targets}, along with the exclusion radii used for each, and other relevant information.

\input{fgas_cluster_table}

\subsection{Data reduction, Spectral Analysis and Non-parametric Deprojection} \label{sec:projct}

The raw \Chandra{} data were cleaned and reduced, and point source masks were created, as described in \morphpaper{}. Blank-field event lists were tailored to each observation, and cleaned in an identical manner. These blank-sky data were renormalized to match the count rates in the science observations in the 9.5--12\keV{} band on a per-CCD basis. Variations in foreground Galactic emission with respect to the blank fields were accounted for, as discussed below.

Clusters centers were determined using 0.6--7.0\keV{}, background-subtracted, flat-fielded images. Initial rough centers were first determined by eye, then centroids were calculated within a radius of 300\,kpc about the initial center (or the largest radius possible without including any of the gaps between CCDs). This centroiding process was iterated a further three times to  ensure convergence. Individual exposures for a given object were checked for consistency, and generally the results from the longest exposure with good spatial coverage were adopted. The final centers were reviewed by eye and slightly adjusted in some cases, e.g.\ due to the presence of asymmetry at small cluster radii, the overall strategy being to choose a center appropriate for the large-scale cluster emission.

\newpage

The spectral analyses described below were all carried out using {\sc xspec}\footnote{\url{http://heasarc.gsfc.nasa.gov/docs/xanadu/xspec/}} (version 12.8.0). Thermal emission from hot, optically thin gas in the clusters, and the local Galactic halo, was modeled as a sum of Bremsstrahlung continuum and line emission components, evaluated using the {\sc apec} plasma model (ATOMDB version 2.0.1). Relative metal abundances were fixed to the solar ratios of \citet{Asplund0909.0948}, with the overall metallicity allowed to vary. Photoelectric absorption by Galactic gas was accounted for using the {\sc phabs} model, employing the cross sections of \citet{Balucinska1992ApJ...400..699B}. For each cluster field, the equivalent absorbing hydrogen column densities, \NH{}, were fixed to the values from the HI survey of \citet{Kalberla0504140}, except for cases where the published values are $>10^{21}\cm^{-2}$ (PKS\,0745 and Abell 478; in these cases, \NH{} was included as a free parameter in our fits, and \tabref~\ref{tab:targets} lists the constraints\footnote{As discussed in \citet{Allen1993MNRAS.262..901A}, the absorption towards Abell 478 varies significantly with projected radius within $\sim5$ arcmin of the cluster center. We address this by fitting the data two ways. First, we perform our usual analysis, allowing \NH{} to vary as a function of radius, finding a declining profile consistent with the results of \citet{Allen1993MNRAS.262..901A}. Second, we perform a fit with a single free value of \NH{}, but we exclude data at energies $<2\keV$ for radii $<500$ arcsec (at radii $>500$ arcsec, our measured \NH{} profile is approximately constant and in agreement with \citealt{Kalberla0504140}). The exclusion of low energies makes the modeled spectra insensitive to the column density. The data are of sufficient quality, and the cluster is hot enough, that the temperature profile can be constrained even excluding this soft band from the analysis over much of the cluster. The two approaches yield consistent mass, temperature and gas density profiles; our reported results are those of the second method.}). The likelihood of spectral models was evaluated using the \citet{Cash1979ApJ...228..939} statistic, as modified by \citet[][the $C$-statistic]{Arnaud1996ASPC..101...17A}. Confidence regions were determined by Markov Chain Monte Carlo (MCMC) explorations of the relevant parameter spaces.

We tested for the possibility of contamination by soft Galactic emission components, over and above that modeled by the blank-sky fields, by analyzing cluster-free regions of the data. This test utilized all regions of the detectors at distances $>r_{200}$ from the cluster center (as estimated from the literature, e.g.\ \arsemf{}; \citealt{Mantz0909.3099, Andersson1006.3068}), provided that at least half of the relevant CCD was included. Spectra in the 0.5--7.0\keV{} band were extracted from each such region, together with appropriate response matrices. Two models were fitted to these spectra: an absorbed power law (photon index $-1.4$), accounting for unresolved AGN, and the same model plus an absorbed, local, solar-metallicity thermal component. The normalizations of both components were permitted to take both positive and negative values. We compared the best-fitting $C$-statistics for the two models using the $F$ distribution, and included a foreground thermal component in subsequent modeling only if the majority of regions tested show evidence for thermal emission at the 95 per cent confidence level. Where regions on different CCDs provided different conclusions, extra weight was given to the CCDs that are better calibrated (e.g.\ chips 0--3 rather than 6) or are more intrinsically sensitive to soft emission (back- rather than front-illuminated). Whenever a foreground model is required, we always fit it simultaneously with other parameters in all subsequent analysis, using the cluster-free data in addition to cluster spectra.

For the minority of nearby clusters where no appropriate cluster-free regions exist in the data, we performed the analysis described below both with and without a foreground component in the model, and discarded the foreground component if its measured normalization was consistent with zero. \tabref~\ref{tab:targets} lists whether a foreground model was required for each cluster.\footnote{Failing to account for an excess foreground component will typically enhance the surface brightness attributed to a cluster and reduce its inferred temperature, with the biases becoming more significant with increasing radius as the true cluster signal falls off. The impact on the \fgas{} values that we ultimately use in this work (measured in a shell spanning 0.8--1.2\,$r_{2500}$; see \secref~\ref{sec:fgasmeas}) depends on a number of factors, including the redshift, temperature and angular extent of each cluster, and the depth of the corresponding observations. Empirically comparing the \fgas{} values derived including and excluding the foreground model for the 22 clusters where our tests find it necessary, we find a bias towards higher \fgas{} values of typically $0.5\sigma$, and as large as $4\sigma$ in the most extreme case. Here $\sigma$ refers to the statistical measurement uncertainty on \fgas{} when the foreground model is erroneously not included. This tends to be slightly smaller than the correct measurement uncertainty.}

Next, we constructed background-subtracted, flat-fielded surface brightness profiles for the clusters in two energy bands: 0.6--2.0\keV{} and 4.0--7.0\keV{}. The soft-band profiles were used to identify radial ranges for the subsequent extraction of spectra in concentric annuli. These annuli were chosen to provide a good sampling of the shape of the brightness profile without the signal being dominated by Poisson fluctuations, with the outermost annulus still containing a clear cluster signal above the background.\footnote{The outermost radius is refined at a later stage, described below. This extra step is particularly necessary when the blank-sky fields do not account for all the non-cluster emission, e.g.\ when a strong Galactic foreground is present.} The hard-band surface brightness profiles were similarly used to define outermost radii where there was clear cluster signal at energies $>4\keV$, a requirement for robustly measuring the temperatures of hot clusters, such as those in our sample. Each cluster thus has three radial ranges defined for it: a central region to be excluded from the mass analysis due to expected dynamical complexity (\secref~\ref{sec:sample}), a shell at intermediate radii where temperatures can be measured robustly, and a shell at large radii where only surface brightness information is used. For each cluster, we generated source spectra and response matrices, and corresponding blank-field background spectra, for the chosen set of annuli. Source spectra were binned to have at least one count in each channel.

We next carried out an initial ``projected'' analysis of the cluster spectra. The cluster emission in each annulus was modeled as an absorbed, redshifted thermal component, with independent normalizations in each annulus but linked temperatures and metallicities. Metal abundances were allowed to vary by a fixed ratio relative to the solar values. For this initial analysis, the temperatures and metallicities were fitted only in the intermediate radial ranges identified for each cluster (i.e.\ excluding the central region and the low signal-to-noise outskirts; see above\footnote{In practice, this was accomplished by creating a duplicate spectrum for each annulus in which the energy range 0.6--2.0\keV{} was binned to a single channel, with other energies ignored. These ``brightness-only'' spectra were used in the central and outer radial ranges, whereas full spectra in the 0.6--7.0\keV{} band were used in the intermediate radial range.}). From these fits, we obtained additional estimates of the foreground model parameters (where applicable) and accurate measurements of the (possibly foreground-subtracted) surface brightness profiles from the normalizations of the cluster components in each of the annuli. Based on these new profiles, we identified and excluded from further analysis any annuli at large radii where the brightness was consistent with zero at 95 per cent confidence, since their inclusion would be problematic for the subsequent spherical deprojection.

\begin{figure}
  \includegraphics[scale=1]{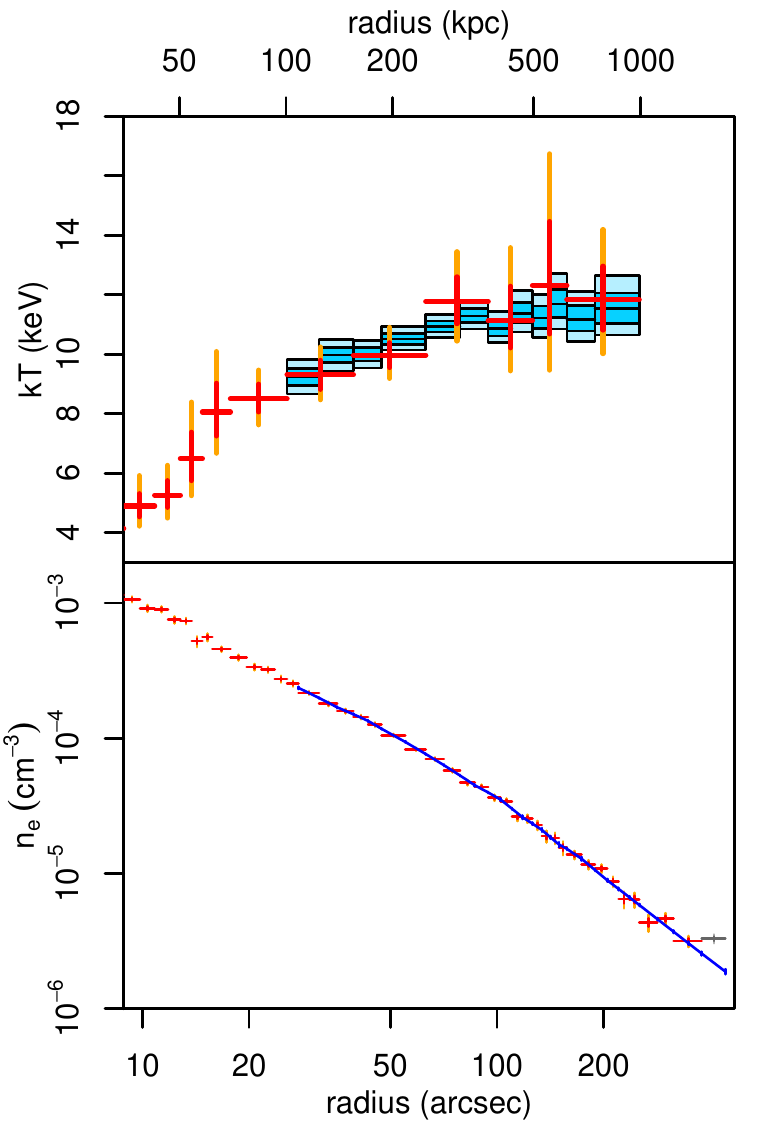}
  \caption{
    Deprojected temperature and electron density profiles for Abell~1835 from our analysis. The normalization of the density profiles is appropriate for our reference cosmology; otherwise, these profiles as a function of angular radius are independent of cosmology (see \appref~\ref{sec:nfwmass}). Blue boxes (top) and lines (bottom) show the results from an analysis which assumes an \NFW{} mass profile and hydrostatic equilibrium and excludes the cluster center. Red/orange crosses show the results of a non-parametric deprojection, including the cluster center. Dark and light colors show the 68 and 95 per cent confidence limits, although note that for the NFW model only the 95 per cent limits are shown in the bottom panel. The agreement of the profiles indicates that the assumption of hydrostatic equilibrium and use of the \NFW{} form of the mass profile provide an acceptable fit to the data. (The disagreement of the outermost, gray point in the density profile is due to the fact that projected emission from larger radii is accounted for in the \NFW{} fit but not in the non-parametric fit.)
  }
  \label{fig:profileeg}
\end{figure}

The data for this refined set of annuli were then fitted with a non-parametric model for the deprojected, spherically symmetric intracluster medium (ICM) density and temperature profiles (the {\sc projct} model in {\sc xspec}). In this model, the cluster atmosphere is described as a set of concentric, spherical shells, with radii corresponding to the set of annuli from which spectra were extracted. Within each shell, the gas is assumed to be isothermal. Given the temperature, metallicity and emissivity (directly related to the density) of the gas in each shell, the spectrum projected onto each annulus can be calculated straightforwardly (e.g.\ \citealt{Kriss1983ApJ...272..439K}). For more details and results based on these fits, including the non-parametric thermodynamic profiles for the clusters, see \profilespaper{}. For the present work, these profiles provide a way to assess the goodness of fit for the \citet[][hereafter \NFW{}]{Navarro9611107} mass model used in determining cluster masses (below). Specifically, the good agreement between the temperature and density profiles obtained under the assumption of an \NFW{} mass profile in hydrostatic equilibrium with the ICM and the non-parametric temperature and density profiles described above (which make no such assumptions) verifies that clusters in our sample are well described by the \NFW{}-hydrostatic equilibrium model (\figref~\ref{fig:profileeg}; similar profiles for all clusters in the sample will be presented in \profilespaper). 

\subsection{Mass and \fgas{} Profile Constraints} \label{sec:nfwfit}

To determine the mass of each cluster, we fit a model that simultaneously describes its three-dimensional mass profile and thermodynamic structure, under the assumptions of spherical symmetry and hydrostatic equilibrium. In this step, we exclude completely data from the central region of each cluster, due to concerns about the validity of these assumptions.\footnote{To be precise, we include annuli from the central region in the {\sc xspec} model, but ignore the corresponding data. Gas temperatures associated with these regions were fixed to broadly reasonable values based on the earlier, non-parametric fits; gas densities are then inferred from these temperatures and the mass profile model. In this way, integrated quantities such as gas masses will provide for the presence of some non-zero amount of gas in the central region, consistent with the remaining model parameters. The gas mass associated with the central region may not be accurate; however, the influence of this exact value on volume-integrated quantities drops rapidly with the outer radius of integration. In particular, the contribution to quantities integrated to $r_{2500}$ is negligible. Note that, in any case, our cosmological analysis uses measurements in a spherical shell that excludes this central region, making these considerations moot for the cosmological results.} Otherwise, the data are used similarly to the projected case, with full spectral information at intermediate radii and only surface brightness at large cluster radii.

The model itself is an adaptation of the {\sc nfwmass} code of \citet[][distributed as part of the {\sc clmass} package for {\sc xspec}; see also \appref~\ref{sec:nfwmass}]{Nulsen1008.2393}.\footnote{The {\sc nfwmass} code contains an option to account for projected emission from spherical radii larger than those otherwise included in the model (i.e.\ beyond the spatial extent of the employed data) by assuming a $\beta$-model continuation of the surface brightness profile. Our sole modification is to set the $\beta$ parameter of this model dynamically, by requiring that the slope of the density (or surface brightness) profile be continuous across this boundary. That is, the value of $\beta$ is set based on the predicted density profile in the outermost shell of the model, itself determined by the mass profile model and the temperature in that shell.} The ICM is again described as a series of concentric, isothermal shells. The mass profile of the cluster is modeled by the \NFW{} form, with two free parameters. Under the assumption of hydrostatic equilibrium, this piecewise-constant temperature profile and \NFW{} mass profile determine the gas density profile up to an overall normalization. (In contrast, the non-parametric model fit in \secref~\ref{sec:projct} allows the temperature and density profiles to be independent, but  without additional assumptions it provides no information about the mass.) We have argued elsewhere \citep{Mantz1106.4052} that ``semi-parametric'' models of the kind used here, combining a non-parametric description of the ICM with a theoretically well motivated, parametrized model for the mass profile, presently provide the least biased approach to X-ray mass determination, given that current data cannot meaningfully constrain non-parametric mass profiles. In contrast, the common assumption of parametrized forms for both the ICM density and temperature profiles represents a complex and non-intuitive prior on the mass profile, and is more constraining than the data require.

A convenient feature of {\sc nfwmass} is that the model itself is completely independent of cosmological assumptions. That is, the fitting procedure described above requires no assumptions about cosmology. The parameter constraints translate to profiles of mass, gas density and temperature (hence also pressure and entropy) of a cluster in physically unmeaningful units, which can be related to physical quantities through a cosmology-dependent factor (see details in \appref~\ref{sec:nfwmass}). By keeping the results in this cosmology-independent form, and by furthermore multiplying the gas density and total mass profiles of each cluster by different, random values when evaluating the results of individual cluster fits, we effectively blinded ourselves to the \fgas{} value of each cluster, the level of agreement among clusters, and any trends with redshift, until the analysis of all clusters was complete and final.

\section{\fgas{} Measurements} \label{sec:fgresults}

The analysis in \secref~\ref{sec:nfwfit} produces temperature, gas density and mass profiles for each cluster, from which gas mass fraction profiles can be derived (see also \appref~\ref{sec:nfwmass}). This section presents those results; for ease of interpretation, these are displayed for a reference \LCDM{} cosmology with $h=0.7$, $\Omegam=0.3$ and $\Omegal=0.7$. Uncertainties for each cluster are based on the distribution of MCMC samples from the spectral analysis, and incorporate the statistical uncertainty in the science observations themselves, the modeling of the astrophysical and instrumental background using the blank-field data, and the constraints on foreground contamination (where applicable).

\subsection{Profiles and Cosmological Measurements} \label{sec:fgasmeas}

\begin{figure*}
  \centering
  \includegraphics[scale=1]{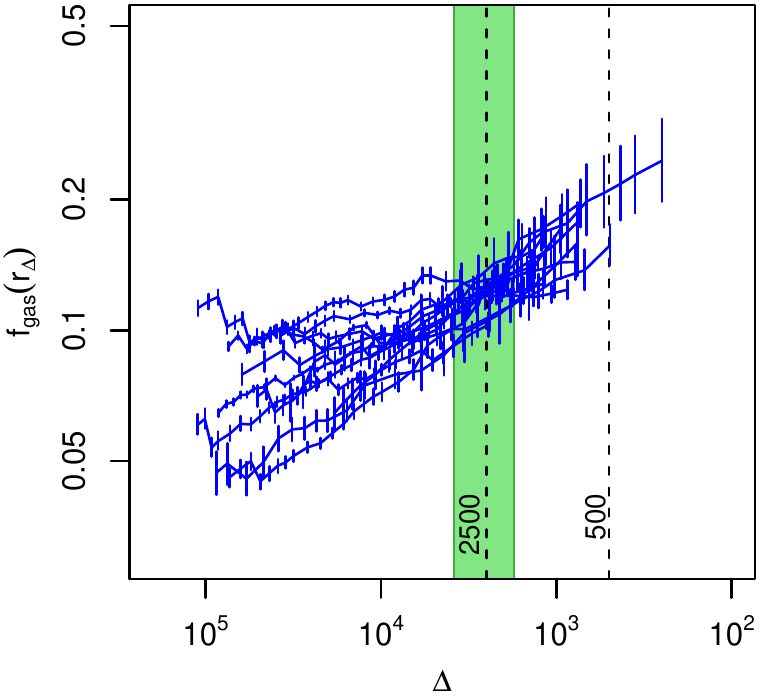}
  \hspace{1cm}
  \includegraphics[scale=1]{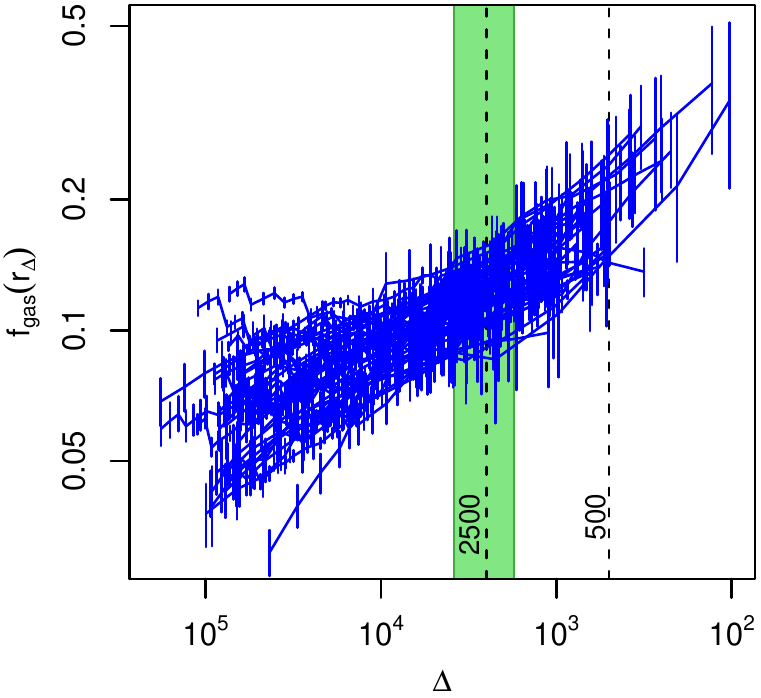}
  \caption{
   Differential \fgas{} profiles as a function of overdensity from our analysis of 13 relaxed clusters at $z \ltsim 0.25$ (left) and all 40 clusters in our sample (right), calculated for our reference cosmology. The shaded region shows the 0.8--1.2\,$r_{2500}$ shell where our cosmological measurements are made (for a typical \NFW{} concentration parameter). The profiles are similar in shape and have small intrinsic scatter at overdensities  $\ltsim 10^4$ ($r \gtsim 0.5\,r_{2500}$). In these figures, we show data for individual clusters only at radii where temperatures were measured. For the few cases where our measurements extend beyond $r_{500}$, we see no evidence of flattening of the profiles.
  }
  \label{fig:fgasdiffindiv}
\end{figure*}

\figref~\ref{fig:fgasdiffindiv} shows the differential \fgas{} profiles (i.e.\ the ratio of gas mass density to total mass density) for the relaxed cluster sample as a function of overdensity, $\Delta=3M/4\pi \rhocr(z) r^3$, where \rhocr{} is the critical density. The left panel of the figure contains the 13 lowest-redshift clusters ($z\ltsim0.25$), while the right panel shows the entire sample. For each cluster, we show results only in the radial range where temperature measurements were performed. While there is greater dispersion at small radii, the profiles largely converge and have small scatter at $\Delta<10^4$ ($r \gtsim 0.5\,r_{2500}$). Outside the cluster centers, the profiles rise with a regular, power-law shape, $\fgas \propto \Delta^{-0.22}$ for $10^4 \geq \Delta \geq 10^3$, or equivalently $\fgas \propto r^{0.43}$ for $0.5 \leq r/r_{2500} \leq 1.6$. At larger radii, fewer than the full sample of 40 clusters provide data; nevertheless the measured profiles remain consistent with this power law, with no indication of flattening. 

\begin{figure*}
  \centering
  \includegraphics[scale=1]{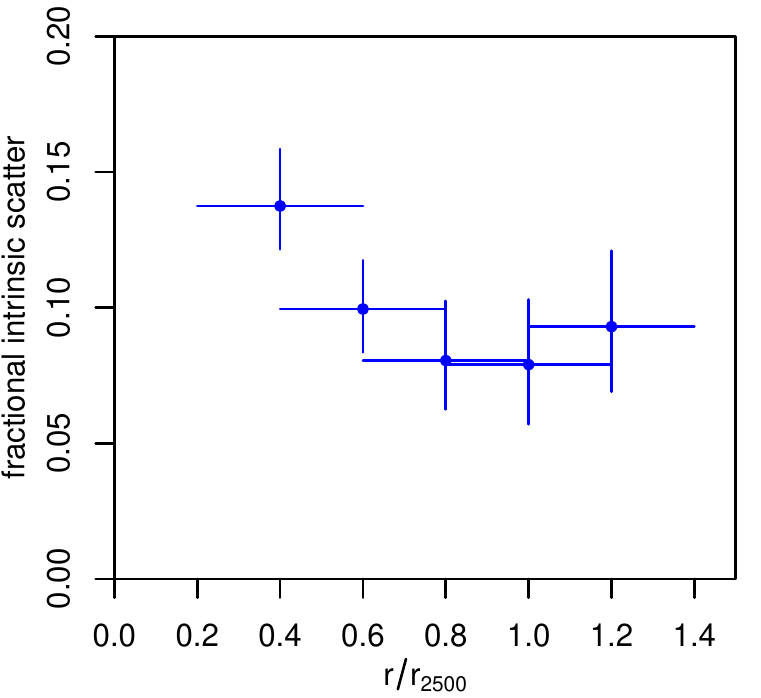}
  \hspace{1cm}
  \includegraphics[scale=1]{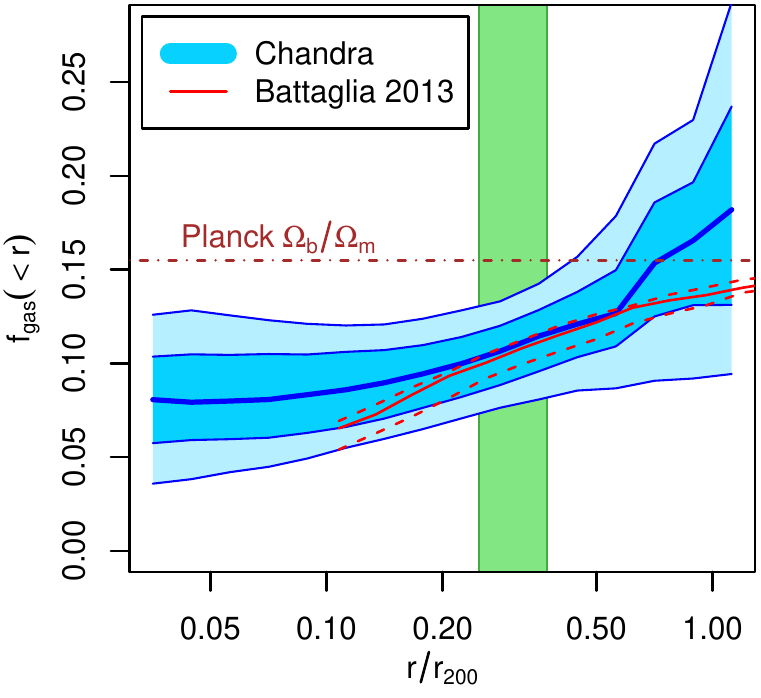}
  \caption{
    Left: The fractional intrinsic scatter of \fgas{} measured in spherical shells (horizontal bars indicate the radial extent of each shell). There is a minimum scatter of 7--8 per cent at radii $\sim r_{2500}$, with a clear increase at smaller radii.
    Right: Cumulative \fgas{} profiles from our analysis of 40 hot, dynamically relaxed clusters compared with the predictions from hydrodynamical simulations. The thick, blue curve shows the median profile observed across our sample, accounting for the measurement uncertainties of each cluster. Dark and light shaded, blue regions show the 68 and 95 per cent confidence limits at each radius, where these probabilities encompass both measurement uncertainties and intrinsic scatter among clusters. As in \figref~\ref{fig:fgasdiffindiv}, each cluster only contributes to the figure at radii where its temperature profile was measured. Red solid and dashed lines show results form the simulations of \citet{Battaglia1209.4082}, for massive ($M_{200}>3\E{14}\Msun$), relaxed clusters, where we have scaled the simulated depletion profile by the cosmic baryon fraction assumed in the simulations. The horizontal, dot-dash line indicates the cosmic baryon fraction measured by \Planck{} \citep{Planck1303.5076}. The green, shaded, vertical band shows the 0.8--1.2\,$r_{2500}$ shell where our cosmological measurements are made (for a typical \NFW{} concentration parameter). The shape of the measured and simulated profiles agree well over a wide range in radii, in particular spanning the radii where our cosmological measurements are made. In both panels, the displayed values of radius and \fgas{} are those appropriate for our reference cosmology (see \appref~\ref{sec:nfwmass}).
  }
  \label{fig:shellscat}
  \label{fig:fgasB13}
\end{figure*}

To investigate the intrinsic scatter as a function of radius, we extracted gas mass fractions for each cluster in a series of spherical shells, spanning radial ranges of width $0.4\,r_{2500}$. The data for each shell were fitted with a linear function of redshift, to approximately marginalize any cosmological signal (\secref~\ref{sec:model}), with the fractional intrinsic scatter as a free parameter. The results of this exercise are shown in the left panel of \figref~\ref{fig:shellscat}, with the scatter minimized at radii $\sim r_{2500}$ and significantly increasing at smaller radii.

\input{fgas_table}

\begin{figure*}
  \centering
  \includegraphics[scale=1]{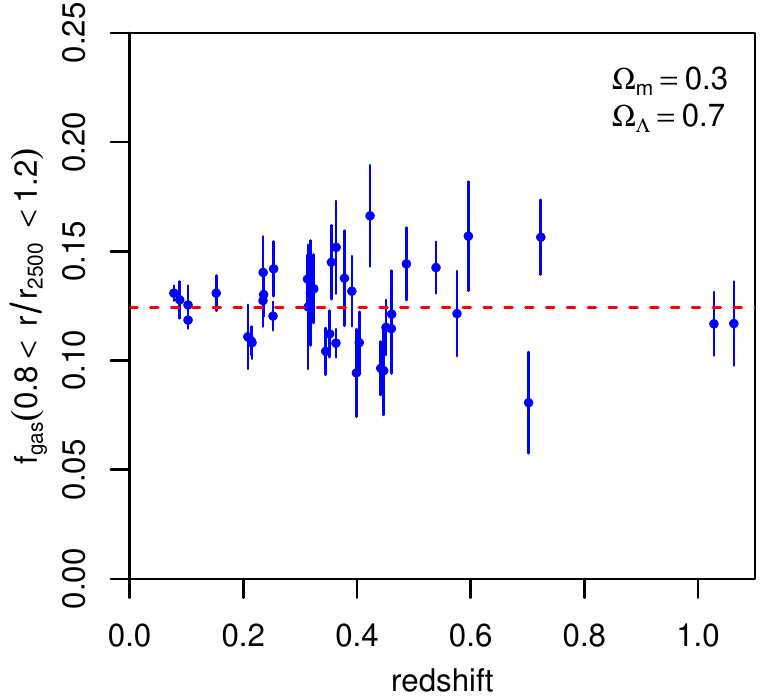}
  \hspace{1cm}
  \includegraphics[scale=1]{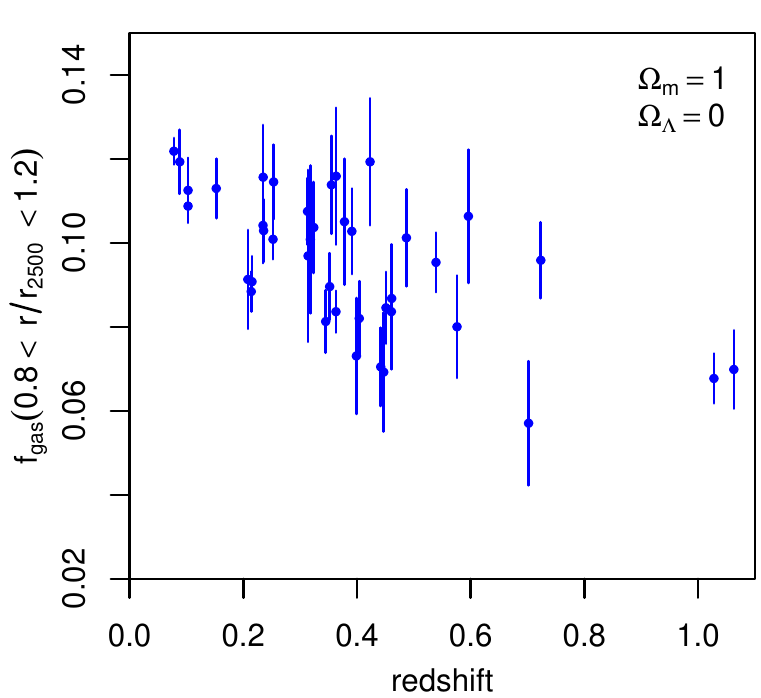}
  \caption{
   Left: Gas mass fractions for our preferred spherical measurement shell about $r_{2500}$ (evaluated for our reference cosmology) are plotted versus redshift. For this cosmology, the data are consistent with a constant value, in agreement with expectations. We address the question of intrinsic scatter in \fgas{} in \secref~\ref{sec:scatter}. Right: \fgas{} values from the same data, derived assuming an SCDM cosmological model with $\Omegam=1$ and without dark energy. The prediction of little or no evolution in \fgas{} (\secref~\ref{sec:fgasmodel}) clearly disfavors this model.
 }
  \label{fig:fgasz}
\end{figure*}
 
The right panel of \figref~\ref{fig:fgasB13} compares our cumulative \fgas{} profiles (i.e.\ integrated within a sphere) to the simulations of \citet{Battaglia1209.4082}. These simulations include the effects of cooling and star formation, as well as heating from AGN feedback, on the ICM, and we specifically plot their results for relatively massive ($3\E{14} < M_{200}/M_\odot < 10^{15}$) and relaxed clusters, where relaxation is defined in terms of the ratio of kinetic to thermal energy. Our measurements are displayed as a dark (light) shaded blue regions, corresponding to 68 (95) per cent confidence at each radius, and representing the combined effect of measurement uncertainties and intrinsic scatter; the thick, blue line is the median $\fgas(<r)$ profile across the cluster sample, again accounting for the measurement uncertainties for each cluster. For context, the horizontal, dot-dashed line shows the cosmic baryon fraction measured by \Planck{} \citep{Planck1303.5076}. We note very good agreement between the shapes of the simulated and measured profiles over a wide range in radius, encompassing the radii of interest for the cosmological measurements, and extending to $\sim r_{500}$ (where our data become increasingly noisy and other astrophysical effects, such as gas clumping, may become important; e.g.\ \citealt{Simionescu1102.2429, Urban1307.3592, Walker1303.4240}).\footnote{Note that the agreement in the normalization of the profiles, while also good, is irrelevant, since the simulations only directly address the depletion parameter, $\Upsilon = \fgas\,\Omegam/\Omegab$. In \figref~\ref{fig:fgasB13}, we have scaled the predicted depletion profile by the cosmic baryon fraction adopted in the simulations.} Note that incompleteness (in the sense that fewer than 40 clusters contribute to the results; see \figref~\ref{fig:fgasdiffindiv}) increases rapidly beyond $\sim r_{1000}$; while it is not clear that selection effects should introduce any particular bias in this case, the combined profile should be treated with caution at large radii.

Our cosmological analysis uses the gas mass fraction integrated within a shell spanning $0.8 < r/r_{2500} < 1.2$, which is shown as a shaded, vertical band in \figref{}s~\ref{fig:fgasdiffindiv} and \ref{fig:fgasB13} (for a typical \NFW{} concentration parameter). The exclusion of smaller radii is intended to minimize both uncertainties in the prediction of the gas depletion factor from hydrodynamic cluster simulations (see \secref~\ref{sec:fgasmodel}) and the intrinsic scatter seen at small radii in the figures, which should result in tighter cosmological constraints. At the same time, temperature profiles (and thus \fgas{}) cannot be reliably measured at radii much larger than $\sim1.6\,r_{2500}$ for most clusters, as can be seen in \figref~\ref{fig:fgasdiffindiv}. In practice, the 0.8--1.2\,$r_{2500}$ shell represents a good compromise between these considerations and the need to maintain good statistical precision of the \fgas{} measurements. \tabref~\ref{tab:fgas} contains our \fgas{} measurements in this shell, along with masses within $r_{2500}$ and redshifts for each cluster. Note that the tabulated \fgas{} values are marginalized over the uncertainty in $r_{2500}$ (or $M_{2500}$,  equivalently).

The behavior with redshift of \fgas{} measured in the 0.8--1.2\,$r_{2500}$ shell (for the adopted reference cosmology with $\Omegam=0.3$ and $\Omegal=0.7$) is shown in the left panel of \figref~\ref{fig:fgasz}. Qualitatively, it is clear that there is little or no evolution with redshift for this cosmological model. The right panel of the figure shows the \fgas{} values derived from the same data, but assuming a cosmology with no dark energy and $\Omegam=1$; for this model, there is an evident redshift dependence. As described more fully in \secref~\ref{sec:model}, this dependence of the \emph{apparent} evolution of $\fgas$ on the cosmic expansion is the basis of dark energy constraints using these data.

\subsection{Mass Dependence} \label{sec:fgasM}

\begin{figure}
  \centering
  \includegraphics[scale=1]{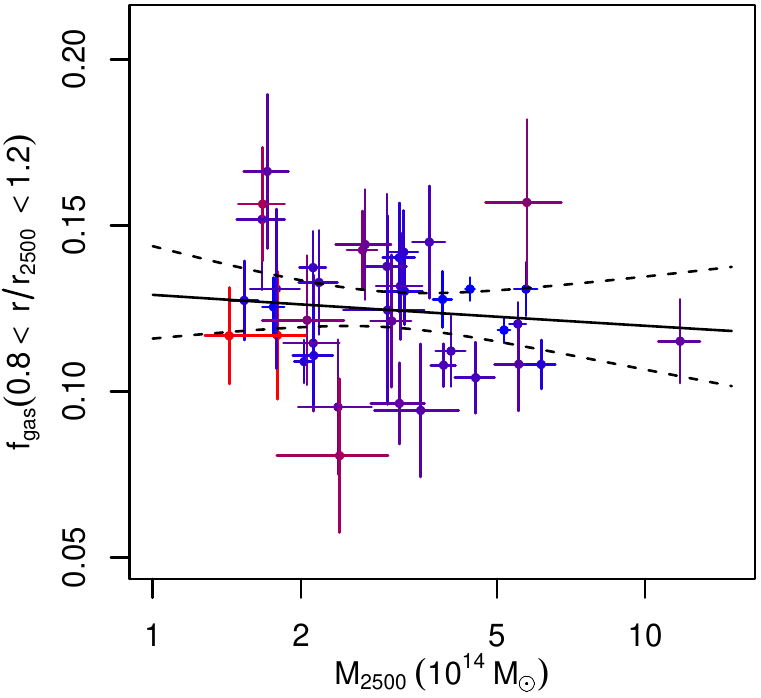}
  \caption{
   Gas mass fractions in our preferred measurement shell (0.8--1.2\,$r_{2500}$, evaluated for our reference cosmology) are plotted versus total mass integrated within $r<r_{2500}$. Lines indicate the best-fitting power law and 95.4 per cent limits, which are consistent with a constant value. The data points are color-coded by redshift (blue to red with increasing $z$; see \tabref~\ref{tab:fgas}).
  }
  \label{fig:fgasM}
\end{figure}
 
Hydrodynamic simulations of cluster formation generally predict a mild increasing trend of the cumulative gas mass fraction, e.g.\ $\fgas(r<r_{2500})$, with mass when fit over a wide mass range extending from group to cluster scales (e.g.\ \citealt{Young1007.0887, Battaglia1209.4082, Planelles1209.5058}, and other references in \secref~\ref{sec:cosintro}). Comparison of \fgas{} values measured for groups and intermediate-mass clusters supports this picture \citep{Sun0805.2320}. It is less clear whether an increasing trend persists at the high masses relevant for this work; $\fgas(r<r_{2500})$ measurements by \citet{Vikhlinin0507092} and \arsemf{} are both consistent with being constant with temperature (hence with mass) for $kT>5\keV$ clusters. We address this question with the current data in \profilespaper{}. Here, we are concerned only with a possible mass trend of \fgas{} integrated in a shell about $r_{2500}$, which has not been studied previously in either simulations or real data.

In \figref~\ref{fig:fgasM}, we show our \fgas{} measurements in the $0.8<r/r_{2500}<1.2$ shell versus $M_{2500}$. Also shown is the best-fitting power-law \fgas--$M$ relation (and 95.4 per cent limits), derived using the Bayesian regression code of \citet{Kelly0705.2774}. Critically, this method accounts for both intrinsic scatter in \fgas{} and the significant anti-correlation between measured values of \fgas{} and $M_{2500}$ (typical correlation coefficients $\sim -0.85$). The best-fitting slope is slightly negative and consistent with zero ($-0.03\pm0.04$; 68.3 per cent confidence limits).\footnote{This question can be investigated in a less cosmology-dependent way by incorporating a power-law mass dependence into the model given below in \secref~\ref{sec:model} and fitting for the slope of this relation simultaneously with the full set of model parameters, again accounting for the anti-correlation between \fgas{} and mass measurements. In this way, uncertainty in the cosmic expansion history can be straightforwardly marginalized over. We obtain consistent results from this analysis, with no evidence for a trend in the shell \fgas{} value. As there is no theoretical motivation for a decreasing trend with mass at radii $\sim r_{2500}$, and since marginalizing over an \fgas--$M$ slope has a negligible effect on our cosmological constraints, we fix the mass dependence to zero in the subsequent sections.}

\subsection{Intrinsic Scatter} \label{sec:scatter}

Thanks to new X-ray observations obtained since \arsemf{}, our data are now precise enough to detect the presence of intrinsic scatter in the \fgas{} measurements. This scatter reflects cluster-to-cluster variations in gas depletion, non-thermal pressure, asphericity, and departures from the \NFW{} mass model. A log-normal scatter in \fgas{}, $\sigmaf$,  is included in the complete model described in \secref~\ref{sec:model} and constrained simultaneously with the rest of the parameters in all our subsequent results. However, constraints on the scatter itself are independent of the cosmological model employed; we find $\sigmaf=0.074\pm0.023$. This 7.4 per cent intrinsic scatter in $\fgas$ corresponds to only $\sim5$ per cent intrinsic scatter in the cosmic distance inferred from a single cluster (\secref~\ref{sec:modelsummary}).

Qualitatively, \figref~\ref{fig:fgasz} appears to show an increase in scatter from $z=0$ to $z\sim0.5$, although the highest redshift points again appear to have little dispersion. Although a trend of \fgas{} scatter with redshift is certainly astrophysically plausible for the cluster population at large, it is not clear that we should expect one for a sample which is restricted to the hottest, most dynamically relaxed clusters at all redshifts. To test for such a trend, we break the data into the redshift ranges 0.0--0.2, 0.2--0.3, 0.3--0.4, 0.4--0.5 and 0.5--1.1, respectively containing 5, 8, 12, 8 and 7 clusters, and fit each subset individually.\footnote{Specifically, we marginalize over non-flat \LCDM{} models with $0<\Omegam<1$, $0<\Omegal<2$ and $0<\fbar<1$ (see \secref~\ref{sec:cosmomodel}). The cosmological parameters are not well constrained by these sub-samples of the data (though see \secref~\ref{sec:lowz}), but this procedure effectively marginalizes over a wide range of plausible cosmic expansion histories within each redshift bin.} The constraints on the intrinsic scatter in each bin agree at $1\sigma$ confidence. Consistently, a weighted linear regression on $\sigmaf(z)$ using these measurements finds no evidence for a non-zero slope with redshift. We henceforth adopt a constant-scatter model throughout this work, while noting that the possibility of evolution will be an interesting question to return to as the number of known high-redshift relaxed clusters continues to grow. 

Observationally, we cannot distinguish between the various possible causes of scatter at this point (though a larger weak lensing/X-ray calibration sample, coupled with ASTRO-H or other X-ray measurements of gas motions, may eventually directly constrain the scatter in non-thermal support), but note that the observed $7.4\pm2.3$ per cent scatter places an upper limit on the individual contributions of the sources mentioned above. This limit is consistent with expectations; for example, the simulations of \citet{Battaglia1209.4082} indicate a fractional scatter of $\sim6$ per cent in the integrated $r<r_{2500}$ gas depletion for massive, relaxed clusters. A similar level of dispersion is expected due to non-thermal pressure \citep{Nagai0609247, Rasia1201.1569, Nelson1308.6589}.

\section{Modeling} \label{sec:model}

This section describes the complete model fitted to the data, including descriptions of both the cosmological expansion and the internal structure of clusters. \tabref~\ref{tab:paramdef} summarizes the parameters of the cluster model and associated priors, as well as the parametrization of the cosmological background used when analyzing cluster or supernova data alone (discussed in more detail below). For completeness, \tabref~\ref{tab:paramcmb} provides the equivalent information for the alternative cosmological parametrization used when analyzing CMB or baryon acoustic oscillation (BAO) data, either alone or in combination with other data sets (this is the standard parametrization in {\sc cosmomc}).

\subsection{Cosmological Model} \label{sec:cosmomodel}

In this paper, we consider cosmological models with a Friedmann-Robertson-Walker metric, containing radiation, baryons, neutrinos, cold dark matter, and dark energy. We adopt an evolving parametrization of the dark energy equation of state \citep{Rapetti0409574},
\begin{equation} \label{eq:wdef}
  w = w_0 + w_a \left( \frac{z}{z+\mysub{z}{tr}} \right) = w_0 + w_a \left( \frac{a^{-1}-1}{a^{-1}+\atr^{-1}-2} \right),
\end{equation}
where $a=(1+z)^{-1}$ is the scale factor. In this model, $w$ takes the value $w_0$ at the present day and $\wet=w_0+w_a$ in the high-redshift limit, with the timing of the transition between the two determined by \atr{}. \eqnref~\ref{eq:wdef} contains as special cases the cosmological constant model (\LCDM{}; $w_0=-1$ and $w_a=0$), constant-$w$ models ($w_a=0$), and the simpler evolving-$w$ model adopted by \citet{Chevallier0009008} and \citet{Linder0208512} ($\atr=0.5$). \arsemf{} provide details on the calculation of cosmic distances using this model.

Beyond the dark energy equation of state, the relevant cosmological parameters for the analysis of cluster data are the Hubble parameter and the present-day densities of baryons, matter, and dark energy. As noted in \appref~\ref{sec:nfwmass}, the interpretation of our X-ray data also depends (extremely weakly) on the primordial mass fraction of helium, \YHe{}. This we derive self-consistently from the baryon density, $\Omegab h^2$, assuming the standard effective number of neutrino species, $\mysub{N}{eff}=3.046$, using the BBN calculations of \citet[][see also \citealt{Hamann0712.2826}]{Pisanti0705.0290}. We note, however, that simply taking $\YHe=0.24$ results in identical cosmological constraints from the \fgas{} data.

\subsection{Gas Depletion} \label{sec:fgasmodel}

Following \arsemf{}, we describe the depletion of X-ray emitting gas in the 0.8--1.2\,$r_{2500}$ shell relative to the cosmic baryon fraction as $\Upsilon(z)= \Upsilon_0(1+\Upsilon_1 z)$, where $\Upsilon_0$ and $\Upsilon_1$ parametrize the normalization and evolution of this quantity. Key differences from previous work are the use of \fgas{} in a shell rather than the cumulative quantity $\fgas(<r_{2500})$, and the fact that we model directly the hot gas depletion rather than both the baryonic depletion and the ratio of mass in stars and cold gas to hot gas. The latter development is due to improvements in hydrodynamical simulations of cluster formation, which now account for a realistic amount of energy feedback from AGN in addition to radiative cooling and star formation. The decision to make our measurements in spherical shells excluding the clusters' centers makes the predictions from simulations yet more reliable.

\begin{figure}
  \centering
  \includegraphics[scale=1]{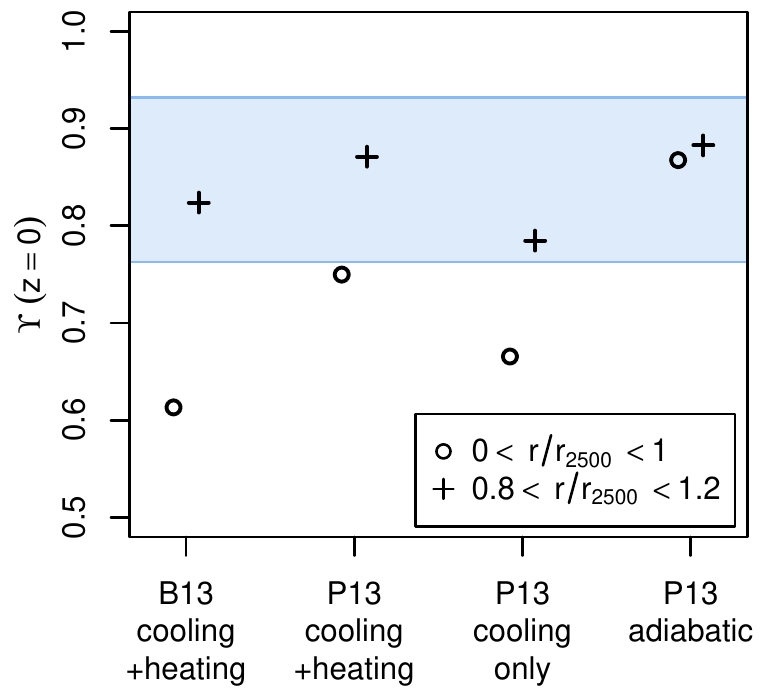}
  \caption{
    Mean gas depletion parameters from the simulations of \citet{Battaglia1209.4082} and \citet{Planelles1209.5058}, integrated in different volumes of the cluster, for simulations including both cooling and heating processes, cooling processes only, or only the most basic ``adiabatic'' gas physics. The agreement among the simulations is substantially improved when considering a spherical shell about $r_{2500}$ rather than the depletion in a complete sphere, including the cluster center. Shading indicates the uniform prior on the $z=0$ depletion that we adopt for the 0.8--1.2\,$r_{2500}$ shell.
  }
  \label{fig:simdepletion}
\end{figure}

Specifically, we consider the recent simulations of \citet{Battaglia1209.4082} and \citet{Planelles1209.5058}, which implement both cooling and AGN feedback in the smoothed particle hydrodynamics (SPH) framework. The $z=0$ gas depletion from these simulations is shown in \figref~\ref{fig:simdepletion}, evaluated both in a sphere of radius $r_{2500}$ and in a spherical shell encompassing $0.8<r/r_{2500}<1.2$. The figure shows that the results of the two independent simulations are in much closer agreement for the spherical shell, excluding the cluster center, than for the full volume. Agreement between the two is at the $\sim5$ per cent level, similar to the level of agreement between these entropy-conserving SPH codes and simulations using adaptive mesh refinement (e.g.\ \citealt{Kravtsov0501227}). On this basis, we adopt a uniform prior on $\Upsilon_0$ centered on $0.848$ (the average of the two cooling+feedback simulation results) and with a full width of 20 per cent, shown by a shaded band in the figure. Note that this conservative prior also encompasses the depletion values derived from the adiabatic and cooling-only simulations of \citet{Planelles1209.5058} for the 0.8--1.2\,$r_{2500}$ shell.

The available information from the published simulations is insufficient to repeat this exercise at $z>0$ to obtain a prior on $\Upsilon_1$ for the shell. However, both works do consider the evolution of the cumulative depletion factor for cooling+feedback models. Neither set of simulations shows evidence for evolution in the gas depletion in massive clusters at the radii of interest (see \figref~10 of \citealt{Battaglia1209.4082} and \figref~7 of \citealt{Planelles1209.5058}). We therefore adopt a conservative uniform prior $-0.05<\Upsilon_1<0.05$.\footnote{\citet{Planelles1209.5058} suggest a prior $-0.02<\Upsilon_1<0.07$ for the {\it baryonic} (not gas) depletion. This range encompasses cumulative results at both $r_{2500}$ and $r_{500}$ for adiabatic and cooling-only simulations in addition to cooling+feedback. Given that the only results in that work that display a trend with redshift apply to the baryonic depletion within $r_{500}$ in simulations without feedback (in particular, the gas depletion is always consistent with zero evolution), we have chosen to adopt a prior whose width is similar to the \citet{Planelles1209.5058} recommendation, but which is centered at zero.}

\begin{table*}
  \centering
 \caption{Parameters and priors used in our analysis of cluster data alone. (When analyzing supernova data alone, we also use this parametrization, though with \fbar{} and the cluster-specific parameters fixed, and without the priors on $h$ and $\Omegab h^2$.) Where no entry appears in the prior column, the prior was uniform and significantly wider than the marginal posterior for that parameter. $\mathcal{N}(\mu,\sigma)$ represents the normal distribution with mean $\mu$ and variance $\sigma^2$, and $\mathcal{U}(x_1,x_2)$ the uniform distribution with endpoints $x_1$ and $x_2$.}
  \vspace{1ex}
  \begin{tabular}{lcll}
    \hline
    Type & Symbol & Meaning & Prior \\
    \hline\vspace{-1.5ex}\\
    Cosmology
    & $h$ & Hubble parameter & $\dnorm(0.738, \, 0.024)$ \\
    & $\fbar$ & Cosmic baryon fraction, $\Omegab/\Omegam$ & \\
    & $\Omegam$ & Total matter density normalized to $\rhocr$ & \\
    & $\Omegade$ & Dark energy density normalized to $\rhocr$ & \\
    & $w_0$ & Present-day dark energy equation of state & \\
    & $w_a$ & Evolution parameter for $w(a)$ & \\
    & $\atr$ & Transition scale factor for $w(a)$ & $\dunif(0.5, \, 0.95)$\vspace{1ex}\\
    Derived
    & $100\,\Omegab h^2$ & Baryon density & $\dnorm(2.202, \, 0.045)$ \\
    & $\YHe$ & Primordial helium mass fraction & \\
    & $\wet$ & Early-time dark energy equation of state &\vspace{1ex}\\    
    Clusters
    & $\Upsilon_0$ & Gas depletion $(\fgas/\fbar)$ normalization & $\dunif(0.763, \, 0.932)$ \\
    & $\Upsilon_1$ & Gas depletion evolution & $\dunif(-0.05, \, 0.05)$ \\
    & $\eta$ & Power-law slope of shell $\fgas$ & $\dnorm(0.442, \, 0.035)$ \\
    & $\sigmaf$ & Intrinsic scatter of shell \fgas{} measurements& \\ 
    & $K_0$ & Mass calibration at $z=0$ & \\
    & $K_1$ & Mass calibration evolution & $\dunif(-0.05,0.05)$ \\
    & $\sigma_K$ & Intrinsic scatter of lensing/X-ray mass ratio & \vspace{1ex}\\
   \hline
  \end{tabular}
  \label{tab:paramdef}
\end{table*}

\begin{table*}
  \centering
  \caption{As \tabref~\ref{tab:paramdef}, but for the cosmological parameters used in joint analysis of cluster \fgas{} and CMB data (as well as CMB alone). We also use this parametrization, with the addition of our standard Gaussian priors on $h$ and $\Omegab h^2$, when analyzing BAO data alone. Neutrino parameters were fixed to the specified values. Note that we do not use priors on $h$  or $\Omegab h^2$ when combining \fgas{} and CMB data. The uniform prior on $h$ below is relevant only for the analysis of CMB data alone. When using CMB data, we also marginalize over the set of nuisance parameters associated with each data set in {\sc cosmomc} (e.g.\ accounting for the thermal Sunyaev-Zel'dovich effect and various astrophysical foregrounds).}
  \vspace{1ex}
  \begin{tabular}{lcll}
    \hline
    Type & Symbol & Meaning & Prior \\
    \hline\vspace{-1.5ex}\\
    Cosmology
    & $\Omegab h^2$ & Baryon density & \\
    & $\mysub{\Omega}{c} h^2$ & Cold dark matter density & \\
    & $\theta_\mathrm{s}$ & Angular size of the sound horizon at last scattering & \\
    & $\Omega_k$ & Effective density from spatial curvature & \\
    & $w_0$ & Present-day dark energy equation of state & \\
    & $w_a$ & Evolution parameter for $w(a)$ & \\
    & $\atr$ & Transition scale factor for $w(a)$ & $\dunif(0.5, \, 0.95)$ \\
    & $\tau$ & Optical depth to reionization & \\
    & $\log\,10^{10}\mysub{A}{s}$ & Scalar power spectrum amplitude & \\
    & $\mysub{n}{s}$ & Scalar spectral index & \\
    & $\Sigma\,m_\nu$ & Species-summed (degenerate) neutrino mass in eV & $=0.056$ \\
    & $\mysub{N}{eff}$ & Effective number of neutrino species & $=3.046$\vspace{1ex}\\
    Derived
    & $h$ & Hubble parameter & $\dunif(0.2, \, 2)$ \\
    & $\YHe$ & Primordial helium mass fraction & \\
    & $\wet$ & Early-time dark energy equation of state &\vspace{1ex}\\    
    \hline
  \end{tabular}
  \label{tab:paramcmb}
\end{table*}

\subsection{Measurement, Calibration and Scatter} \label{sec:miscmodel}

Any inaccuracies in instrument calibration, as well as any bias in measured masses due to substructure, bulk motions and/or non-thermal pressure in the cluster gas, will cause the measured values of \fgas{} to depart from the true values. With the advent of robust gravitational lensing measurements (\wtg{}), these effects can now be directly constrained from data.\footnote{Strictly speaking, the lensing data can only calibrate bias in the X-ray mass determinations, not any bias in gas masses. However, the current level of uncertainty in total mass, $\sim10$ per cent, is significantly greater than the systematic uncertainty in the flux calibration of \Chandra{} (for example, taking the level of disagreement between the ACIS and XMM-{\it Newton} detectors as the scale of the uncertainty). The lensing mass measurements themselves are expected to be unbiased (see \citealt{Becker1011.1681} and \wtg{}).} From the 12 clusters in common between this work and the \wtg{} sample, we (Applegate et~al., in preparation) find a mean weak lensing to \Chandra{} X-ray mass ratio of $K=0.90\pm0.09$ for our reference cosmology.\footnote{This analysis incorporates allowances for systematic uncertainties, as detailed in \wtg{}. In particular, systematics associated with galaxy shear measurements, photometry and projection are individually controlled at the few per cent level.}$^,$\footnote{Note that an underestimate of the total mass by the X-ray analysis, as one might expect due to non-thermal support (e.g.\ \citealt{Nagai0609247}), would correspond to values $K>1$. The measurement of $K<1$ (albeit at a relatively low confidence level) implies that temperature measurements based on fitting the Bremsstrahlung continuum to \Chandra{} observations (with the current calibration) may be overestimated by $\gtsim10$ per cent at the typical temperatures of our cluster sample (5--12\keV{}). This estimate would place the ``correct'' temperatures roughly midway between \Chandra{} ACIS and XMM-{\it Newton} MOS results from continuum fitting, and in broad agreement with results from fitting the Fe emission line with either instrument (8th IACHEC meeting; \url{http://web.mit.edu/iachec/meetings/2013/index.html}). See Applegate et~al. (in preparation) for more details.}

This constraint has a mild dependence on the cosmological background, due to the dependence of the lensing signal on angular diameter distances. Rather than taking the above result as a prior, therefore, we directly incorporate the data and analysis used by Applegate et al.\ into our model (see that work for details of the gravitational lensing likelihood calculation). Specifically, we model the mean ratio of lensing to X-ray mass as $K(z)=K_0(1+K_1 z)$, with a log-normal intrinsic scatter, and constrain these parameters simultaneously with the rest of the model. The evolution parameter, $K_1$, cannot be constrained by the 12 clusters in the calibration sub-sample; while there is no particular theoretical expectation for evolution in, e.g., the amount of non-thermal pressure in the most relaxed clusters, we nevertheless marginalize over a uniform prior $-0.05<K_1<0.05$.

Additionally, we must account for the fact that our X-ray measurements are made under the assumption of a particular reference cosmological model. The tabulated \fgas{} values are thus proportional to $[d^\mathrm{ref}(z)/d(z)]^{3/2}$, where $d(z)$ is the true cosmic distance to the cluster, and $d^\mathrm{ref}(z)$ is the distance evaluated assuming the reference model.\footnote{We do not distinguish between angular diameter and luminosity distances in this section, but see \appref~\ref{sec:nfwmass}.} Another, smaller dependence arises through the dependence of the reference value of $r_{2500}$ (actually the equivalent angular radius, $\theta_{2500}$) on the critical density, $\rhocr(z)$. For a given trial cosmology, we need to predict the gas mass fraction in the reference measurement shell rather than the true 0.8--1.2\,$r_{2500}$ shell (according to the trial cosmology's $\rhocr$). As in \arsemf{}, we take advantage of the fact that the \fgas{} profiles of our clusters are consistent with a simple power law at the relevant radii (\figref~\ref{fig:fgasdiffindiv}). For each cluster, we fit a power-law model to the function $\fgas(0.8x<r/r^\mathrm{ref}_{2500}<1.2x)$, as $x$ varies from 0.7 to 1.3; averaging over the cluster sample, we find a power-law slope of $\eta=0.442\pm0.035$. 

Including these terms, the complete model which we fit to the data is
\begin{eqnarray} \label{eq:fgasmodel}
  \fgas^\mathrm{ref}\left(0.8<\frac{r}{r^\mathrm{ref}_{2500}}<1.2; ~z\right) \hspace{3.4cm}\nonumber\\
  \hspace{4em}= K(z) \, A\, \Upsilon_0(1+\Upsilon_1 z) \left( \frac{\Omegab}{\Omegam} \right) \left[ \frac{d^\mathrm{ref}(z)}{d(z)} \right]^{3/2},
\end{eqnarray}
where (\arsemf)
\begin{equation} \label{eq:thetarat}
  A = \left( \frac{\theta^\mathrm{ref}_{2500}}{\theta_{2500}} \right)^\eta \approx \left( \frac{H(z) \, d(z)}{\left[H(z) \, d(z)\right]^\mathrm{ref}} \right)^\eta.
\end{equation}

\eqnref~\ref{eq:fgasmodel} represents the predicted mean for each of our cluster measurements. In addition, we model and fit for a log-normal intrinsic scatter in the measured value, $\sigmaf$, as described in \secref~\ref{sec:scatter}. The measurement errors of $\fgas^\mathrm{ref}$, after marginalizing over $r^\mathrm{ref}_{2500}$, are approximately log-normal as well, and we model them as such. The likelihood associated with each cluster thus has a simple, Gaussian form, with mean given by the logarithm of \eqnref~\ref{eq:fgasmodel} and variance equal to the sum of $\sigma^2_f$ and the square of the associated fractional measurement error.\footnote{The log-normal form was chosen for computational convenience. However, we have explicitly verified that our cosmological constraints are unchanged if the intrinsic scatter and measurement errors are instead modeled as Gaussian. The residuals from the best fit are consistent with either hypothesis, reflecting the fact that the two distributions are similar for small values of the total fractional scatter.}

\subsection{Summary of the Model and Priors} \label{sec:modelsummary}

Along with the intrinsic scatter in \fgas{}, \eqnref~\ref{eq:fgasmodel} constitutes a complete model for the X-ray \fgas{} measurements. The normalization of this function depends on the product $h^{3/2}\Omegab/\Omegam$, and is systematically limited by the nuisance parameters $K_0$ and $\Upsilon_0$. In practice, the calibration parameter, $K_0$, dominates the error budget, with the statistical uncertainty on the mean \fgas{} value, especially at low redshift, being small. \secref~\ref{sec:lowz} outlines the constraints on cosmological parameters obtained from the low-redshift clusters, for which uncertainties related to the model of dark energy and the evolution of the depletion factor ($\Upsilon_1$) are negligible. In particular, combining the low-redshift cluster data with priors on $h$ and $\Omegab h^2$ produces a tight constraint on $\Omegam$ which is independent of the cosmic expansion.

\begin{figure}
  \centering
  \includegraphics[scale=1]{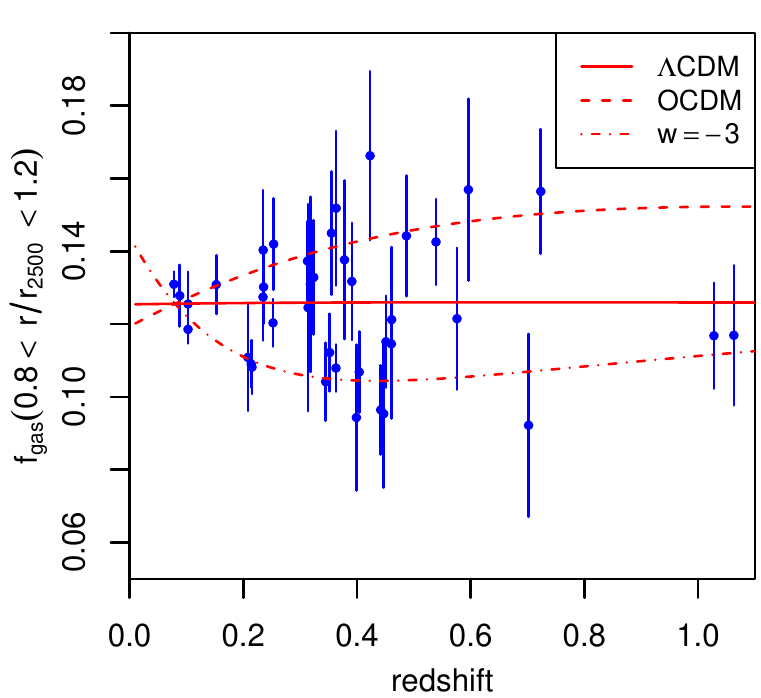}
  \caption{
   The \fgas{} data, measured in the 0.8--1.2\,$r_{2500}$ shell for our reference cosmology, are compared with the predictions of three dark energy models. These model predictions incorporate the full level of detail in \eqnref~\ref{eq:fgasmodel}, i.e.\ they are predictions for exactly what we would measure given the adopted reference cosmology (and for nominal values for the nuisance parameters). Each model prediction is normalized to agree with the data at $z<0.16$, which in practice would constrain the value of $\Omegam$ on their own. The figure thus illustrates the redshift-dependent signal available to the $\fgas(z)$ data once the $\Omegam$ constraint from the normalization of $\fgas$ is accounted for. The solid line shows predictions for a flat \LCDM{} model ($\Omegam=0.3$, $\Omegal=0.7$, $w=-1$; identical to the reference), the dashed line an open model ($\Omegam=0.3$, $\Omegal=0.0$), and the dot-dashed line a flat, constant-$w$ model ($\Omegam=0.3$, $\Omegade=0.7$, $w=-3$).
  }
  \label{fig:hubblediag}
\end{figure}

The redshift dependence of $\fgas^\mathrm{ref}(z)$ provides constraints on dark energy, through the $d(z)^{3/2}$ dependence. This is illustrated in \figref~\ref{fig:hubblediag}, which shows our data along with the predictions (from \eqnref~\ref{eq:fgasmodel}) of three dark energy models. The normalizations of the model curves have been fitted to the $z<0.16$ cluster data to demonstrate the difference between models that might be acceptable to those low-redshift data alone.\footnote{Note that this normalization effectively measures \Omegam{}, as described above. Hence, it is instructive to compare the $d(z)$ curves for various models of dark energy but with the same value of \Omegam{}, as in the figure.} Our sensitivity to the redshift-dependent signal is limited by the systematic uncertainty represented by $\Upsilon_1$ and $K_1$, and the sparsity of data at redshifts $z \gtsim 0.5$; in practice, the latter dominates the uncertainty on dark energy parameters from current data (see also \secref~\ref{sec:future}).

\subsection{Fitting the Models}

The cluster model described in the preceding sections, and the associated likelihood evaluation, have been coded into a stand-alone library that can straightforwardly be linked to {\sc cosmomc}\footnote{\url{http://cosmologist.info/cosmomc/}} or other software.\footnote{\url{http://www.slac.stanford.edu/~amantz/work/fgas14/}} The results presented here were produced using {\sc cosmomc} (\citealt{Lewis0205436}; October 2013 version). Cosmological calculations were evaluated using the {\sc camb} package of \citet*{Lewis9911177}, suitably modified to implement the evolving-$w$ model of \citet{Rapetti0409574}, including the corresponding dark energy density perturbations.\footnote{To calculate the dark energy perturbations in evolving-$w$ models, we do not use the standard Parametrized Post-Friedmann (PPF) framework in {\sc cosmomc}, but rather an extension of the fluid description used for constant-$w$ models. Especially for cases far from \LCDM{}, this gives us more accurate results by construction. We have verified that the prescription we use to avoid the divergence at the crossing of the phantom divide ($w=-1$) allows us to appropriately match the PPF results designed to overcome that theoretical problem \citep*{Fang0808.3125}.}

In \secref~\ref{sec:cosmores}, we compare and combine our \fgas{} cosmological constraints with those of other cosmological probes. Specifically, we include all-sky CMB data from the Wilkinson Microwave Anisotropy Probe (WMAP 9-year release; \citealt{Bennett1212.5225, Hinshaw1212.5226}) and the \Planck{} satellite (1-year release, including WMAP polarization data; \citealt{Planck1303.5075}), as well as high-multipole data from the Atacama Cosmology Telescope (ACT; \citealt{Das1301.1037}) and the South Pole Telescope (SPT; \citealt{Keisler1105.3182, Reichardt1111.0932, Story1210.7231}). For these data, we use the likelihood codes provided by the WMAP\footnote{\url{http://lambda.gsfc.nasa.gov}} (December 2012 version) and \Planck{}\footnote{\url{http://pla.esac.esa.int/pla/aio/planckProducts.html}} teams, where the latter also evaluates the ACT and SPT likelihoods. When using CMB data, we marginalize over the default set of nuisance parameters associated with each data set in {\sc cosmomc} (e.g.\ accounting for the thermal Sunyaev-Zel'dovich effect and various astrophysical foregrounds). In addition, we include the Union 2.1 compilation of type Ia supernovae \citep{Suzuki1105.3470} and BAO data from the combination of results from the 6-degree Field Galaxy Survey (6dFGS; $z=0.106$; \citealt{Beutler1106.3366}) and the Sloan Digital Sky Survey ($z=0.35$ and $0.57$; \citealt{Padmanabhan1202.0090, Anderson1303.4666}). For these data sets, likelihood functions are included as part of {\sc cosmomc}.

\section{Cosmological Results} \label{sec:cosmores}

This section presents the cosmological constraints obtained from our analysis of the cluster data. \secref~\ref{sec:lowz} discusses the constraints available from the lowest redshift clusters, with minimal external priors. The subsequent sections explore progressively more complex cosmological models using the cluster data, as well as independent cosmological probes. When combining data sets, we consider separately combinations which include WMAP or \Planck{} CMB data. For simplicity, the figures and discussion in this section refer to the WMAP version of these results. The combined results using \Planck{} data are quantitatively similar; for completeness we include the corresponding figures in \appref~\ref{sec:planckres}. Our results are summarized in \tabref{}s~\ref{tab:life} and \ref{tab:death}.

\begin{table}
  \begin{center}
    \caption{
      Marginalized best-fitting values and 68.3 per cent maximum-likelihood confidence intervals on cosmological parameters from our analysis of low-redshift ($z<0.16$) clusters, including systematic uncertainties (\secref~\ref{sec:lowz}). Parameters are defined in \secref~\ref{sec:cosmomodel}. These constraints are essentially identical in all cosmological models considered in this work except those with $\atr$ free (i.e. where the dark energy equation of state can vary rapidly at $z<0.16$). Columns 1--2 indicate whether standard priors on $h$ and $\Omegab h^2$ (\tabref~\ref{tab:paramdef}; \citealt{Riess1103.2976, Cooke1308.3240}) are used in addition to the \fgas{} data.
    }
    \label{tab:life}
    \vspace{1ex}
    \begin{tabular}{ccr@{ $=$ }r@{ $\pm$ }l}
      \hline
      \multicolumn{2}{c}{Prior} & \multicolumn{3}{c}{Constraint} \\
      $h$ & $\Omegab h^2$ & \multicolumn{3}{c}{} \\
      \hline\vspace{-1.5ex}\\
      & & $h^{3/2}\Omegab/\Omegam$ & $0.089$ & $0.012$\vspace{1ex}\\
      $\surd$ & & $\Omegab/\Omegam$ & $0.14$ & $0.02$\vspace{1ex}\\
      & $\surd$ & $\Omegam h^{1/2}$ & $0.24$ & $0.03$\vspace{1ex}\\
      $\surd$ & $\surd$ & $\Omegam$ & $0.27$ & $0.04$\vspace{1ex}\\
      \hline
    \end{tabular}
  \end{center}
\end{table}

\subsection{Dark Energy-Independent Constraints from Low-Redshift Data} \label{sec:lowz}

The amount and nature of dark energy have a very small effect on cosmic expansion at the lowest redshifts in our data set, in particular for the 5 clusters with $0.07<z<0.16$.\footnote{We have explicitly verified that our cluster results in this section are identical whether we marginalize over \LCDM{} or flat, constant-$w$ models. This insensitivity is not absolute; for example, it breaks down if the dark energy equation of state is allowed to evolve rapidly at redshifts $z<0.16$, as in our most general dark energy model.} To the extent that the cosmology-dependent curvature of $d(z)$ and the variation of $\Upsilon(z)$ are negligible over this redshift range, \eqnref~\ref{eq:fgasmodel} reduces to
\begin{equation} \label{eq:lowz}
  \fgas^\mathrm{ref} \propto K_0 \, \Upsilon_0 \, \frac{\Omegab}{\Omegam} \, h^{3/2}.
\end{equation}
As our data are very precise for these nearby clusters, constraints on the product $h^{3/2}\,\Omegab/\Omegam$ will be systematically limited, specifically by the calibration parameter, $K_0$ (\tabref~\ref{tab:paramdef}). We obtain $h^{3/2}\,\Omegab/\Omegam = 0.089\pm0.012$.\footnote{Note that this result marginalizes over the complete model; the simplified form in \eqnref~\ref{eq:lowz} is for illustration only.}

\begin{figure}
  \centering
  \includegraphics[scale=1]{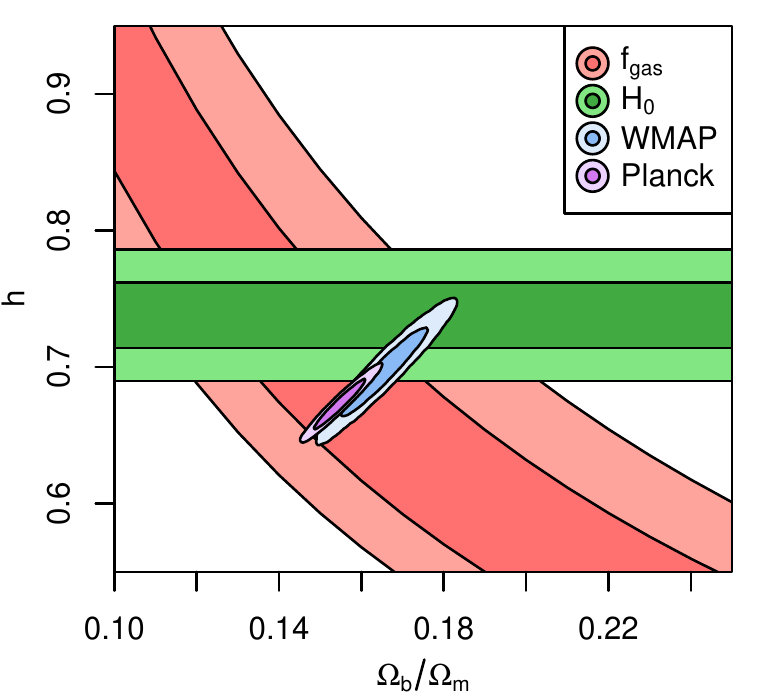}
  \caption{
    Constraints on the Hubble parameter and cosmic baryon fraction from \fgas{} data (red; $z<0.16$ data only), WMAP CMB data (blue; \citealt{Hinshaw1212.5226}), \Planck{} CMB data (purple; \citealt{Planck1303.5076}), and direct measurements of the Hubble expansion (green; \citealt{Riess1103.2976}). Dark and light shaded regions show the marginalized 68.3 and 95.4 per cent confidence regions, accounting for systematic uncertainties. A flat \LCDM{} cosmology is assumed, although the \fgas{} and Hubble expansion data are insensitive to this prior.
  }
  \label{fig:lowzfbh}
\end{figure}

\figref~\ref{fig:lowzfbh} shows this constraint from cluster \fgas{} in the $\Omegab/\Omegam$--$h$ plane, along with measurements of the local Hubble expansion \citep{Riess1103.2976} and the tight constraints for flat \LCDM{} models from WMAP and \Planck{}. The \fgas{} data are consistent with all of these data individually, although the figure shows clearly the tension in the value of $h$ derived from \Planck{} compared with that from the local distance ladder ($h \approx 0.74$). Combining the WMAP and \fgas{} data for flat \LCDM{} models, we obtain a constraint on the Hubble parameter, $h=0.690\pm0.017$, consistent with \Planck{}.

The cluster constraint on $h^{3/2}\,\Omegab/\Omegam$ can be combined with direct Hubble parameter measurements of \citet{Riess1103.2976} to obtain a CMB-free constraint on the cosmic baryon fraction. Applying their constraint of $h=0.738\pm0.024$, we find $\Omegab/\Omegam=0.14\pm0.02$, consistent with the best-fitting WMAP-only and \Planck{}-only values at the $2\sigma$ and $1\sigma$ levels, respectively.\footnote{Note that using instead the results of the Carnegie Hubble Project, $h=0.742\pm0.021$ \citep{Freedman1208.3281}, shifts this constraint by $<1$ per cent. When we additionally use a prior on $\Omegab h^2$, below and in subsequent sections, the influence of $h$ is even smaller (the residual dependence being $h^{-1/2}$; see \eqnref~\ref{eq:lowz}). The effect on dark energy constraints in later sections is completely negligible.} Alternatively, using a prior on $\Omegab h^2$ from BBN data allows the low-$z$ clusters to constrain the combination $\Omegam h^{1/2}$. We employ a prior $100\,\Omegab h^2=2.202\pm0.045$ based on the deuterium abundance measurements of \citet{Cooke1308.3240}, which yields $\Omegam h^{1/2}=0.24\pm0.03$. Combining priors on both $h$ and $\Omegab h^2$ with our measurement of $h^{3/2}\,\Omegab/\Omegam$  provides a direct constraint on \Omegam{} (e.g.\ \citealt{White1993Natur.366..429W}). We find $\Omegam=0.27\pm0.04$ from the $z<0.16$ clusters, in good agreement with the full \fgas{} data set (below), as well as the combination of CMB data with other probes of cosmic distance (e.g.\ \citealt{Hinshaw1212.5226, Planck1303.5076}).

The above priors on $h$ and $\Omegab h^2$ constitute the ``standard'' priors that we use together with the cluster \fgas{} data in subsequent sections (\tabref~\ref{tab:paramdef}).
In models where the equation of state of dark energy is a free parameter, CMB data provide a relatively weak upper bound on $h$. However, because the CMB still tightly constrains $\Omegab/\Omegam$ in this case, the combination of CMB and \fgas{} data provides tight constraints on both $h$ and $\Omegab/\Omegam$ (see also \arsemf). Consequently, we do not require or use the priors on $h$ and $\Omegab h^2$ in later sections where the \fgas{} data are used in combination with CMB measurements.

\subsection{Constraints on \LCDM{} Models}

\begin{figure}
  \centering
  \includegraphics[scale=1]{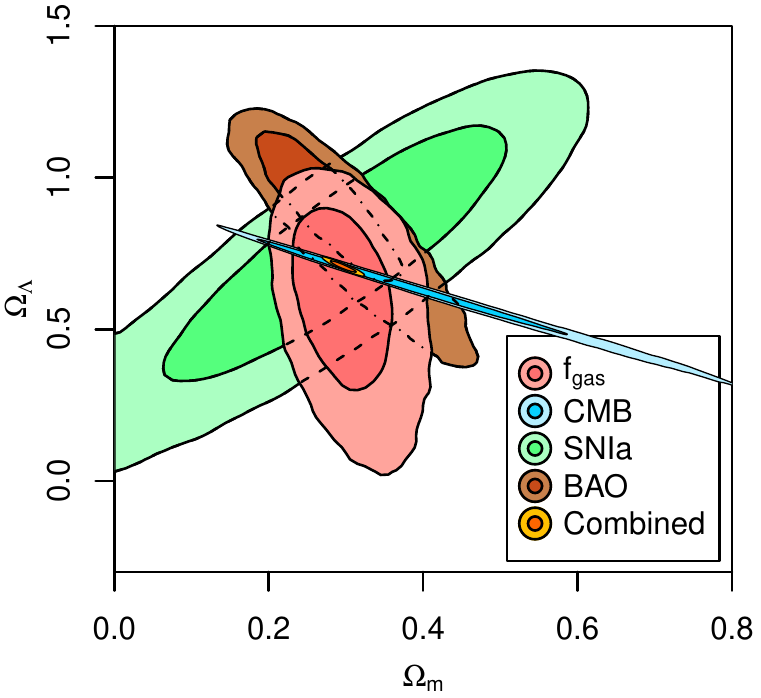}
  \caption{
    Constraints on \LCDM{} models from the full cluster \fgas{} data set (red, including standard priors on $h$ and $\Omegab h^2$), CMB data from WMAP, ACT and SPT (blue; \citealt{Keisler1105.3182, Hinshaw1212.5226, Reichardt1111.0932, Story1210.7231, Das1301.1037}), type Ia supernovae (green; \citealt{Suzuki1105.3470}), baryon acoustic oscillations (brown, also including priors on $h$ and $\Omegab h^2$; \citealt{Beutler1106.3366, Padmanabhan1202.0090, Anderson1303.4666}), and the combination of all four (gold). Dark and light shaded regions show the marginalized 68.3 and 95.4 per cent confidence regions, accounting for systematic uncertainties. Priors on $h$ and $\Omegab h^2$ are not included in the combined constraints.
  }
  \label{fig:lcdm}
\end{figure}

For non-flat \LCDM{} models, the constraints obtained from the full \fgas{} data set (plus standard priors) are shown as red contours in \figref~\ref{fig:lcdm}. We obtain $\Omegam=0.29\pm0.04$ and $\Omegal=0.65^{+0.17}_{-0.22}$, with relatively little correlation between the two parameters, as can be seen in the figure. Also shown in \figref~\ref{fig:lcdm} are independent constraints from WMAP+ACT+SPT (hereafter CMB; \citealt{Keisler1105.3182, Hinshaw1212.5226, Reichardt1111.0932, Story1210.7231, Das1301.1037}), type Ia supernovae \citep{Suzuki1105.3470} and BAO (\citealt{Beutler1106.3366, Padmanabhan1202.0090, Anderson1303.4666}), where the latter constraints also incorporate our standard priors on $h$ and $\Omegab h^2$. The four independent data sets are in good agreement. Combining them (without additional priors), we obtain tight constraints strongly preferring a flat universe: $\Omegam=0.296\pm0.011$ and $\Omegal=0.706\pm0.013$ individually, with $\Omega_k=-0.003\pm0.004$.

\subsection{Constraints on Constant-$w$ Models}

\begin{figure*}
  \centering
  \includegraphics[scale=1]{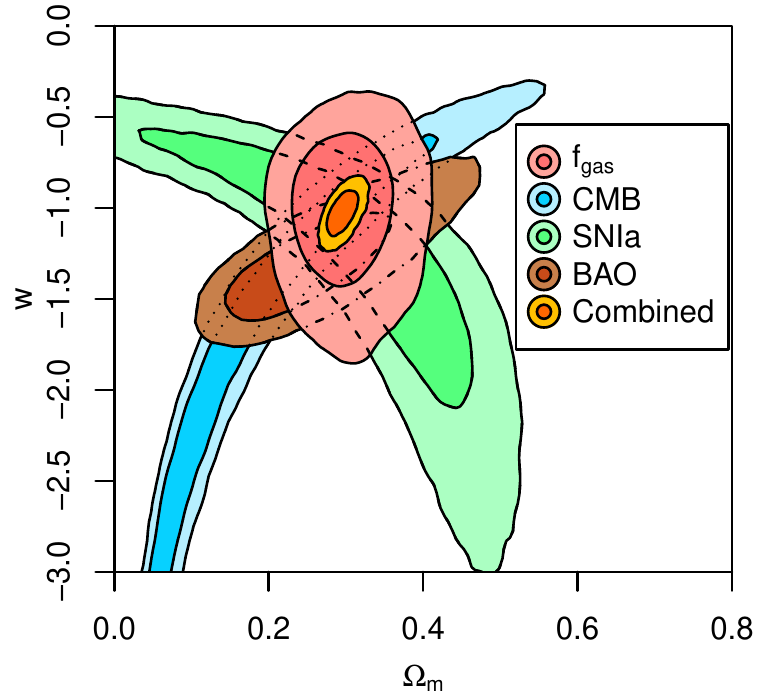}
  \hspace{1cm}
  \includegraphics[scale=1]{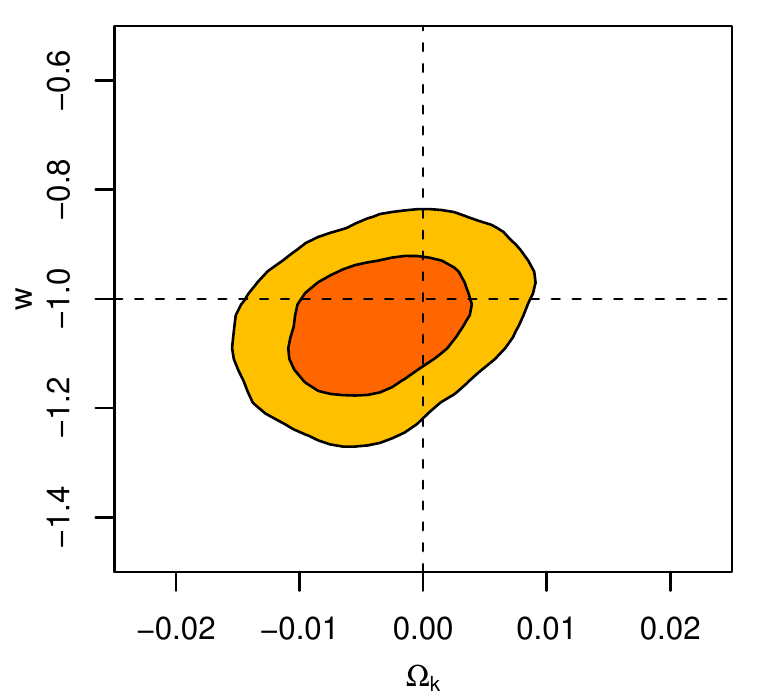}
  \caption{Cosmological constraints from cluster \fgas{} data (red, including standard priors on $h$ and $\Omegab h^2$), CMB data from WMAP, ACT and SPT (blue), type Ia supernovae (green), baryon acoustic oscillations (brown, also including priors on $h$ and $\Omegab h^2$), and the combination of all four (gold). The priors on $h$ and $\Omegab h^2$ are not included in the combined constraints. Dark and light shaded regions show the marginalized 68.3 and 95.4 per cent confidence regions, accounting for systematic uncertainties.
    Left: flat models with a constant dark energy equation of state, $w$. Right: Constraints on non-flat models with constant $w$ from the above combination of data. Vertical and horizontal dashed lines respectively indicate spatially flat models and cosmological-constant models.
 }
 \label{fig:constw}
 \label{fig:constwcurve}
\end{figure*}

We next consider spatially flat models with a constant dark energy equation of state, $w$. The \fgas{} constraint on $\Omegam$ is $0.29\pm0.04$, identical to the \LCDM{} case. Our constraint on the equation of state is $w=-0.98\pm0.26$. The \fgas{} constraints appear in the left panel of \figref~\ref{fig:constw} along with independent constraints from CMB, supernova and BAO data, and the combination of all four. Again, the different cosmological probes are in good agreement; from the combination we obtain $\Omegam=0.296\pm0.014$ and $w=-1.02\pm0.08$.

Allowing global spatial curvature in the model, the combination of \fgas{}, CMB, supernova and BAO data yields $\Omega_k=-0.004\pm0.005$ and $w=-1.04\pm0.08$ (right panel of \figref~\ref{fig:constwcurve}), again consistent with the flat \LCDM{} model.

\subsection{Constraints on Evolving-$w$ Models}

\begin{figure*}
  \centering
  \includegraphics[scale=1]{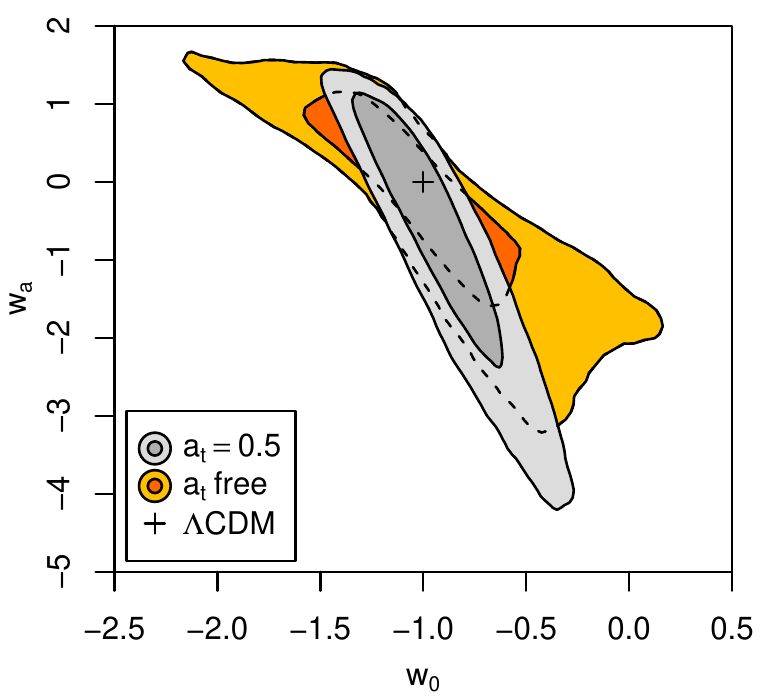}
  \hspace{1cm}
  \includegraphics[scale=1]{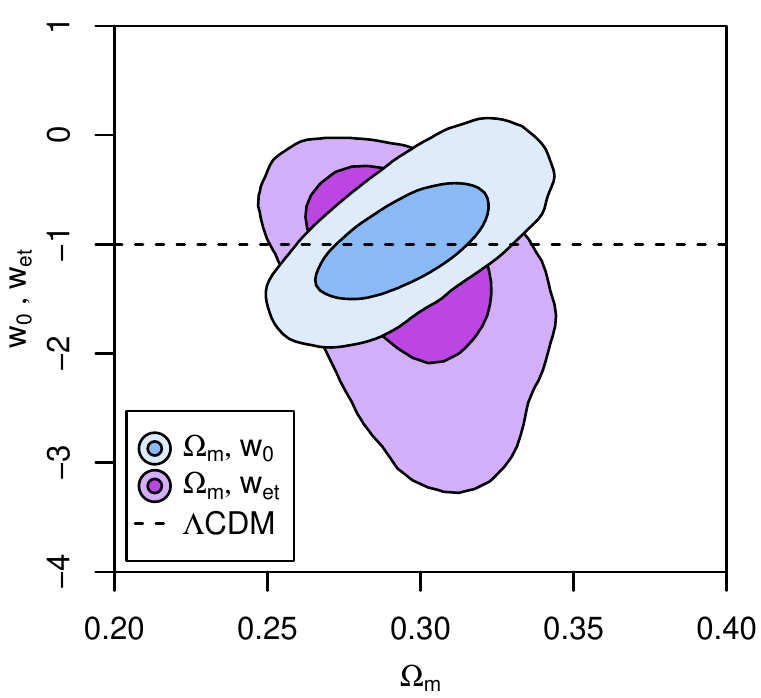}
  \caption{
    Left: Constraints on the present-day dark energy equation of state and its evolution from the combination of cluster \fgas{}, CMB, type Ia supernova, and baryon acoustic oscillation data. Dark and light shaded regions show the marginalized 68.3 and 95.4 per cent confidence regions, accounting for systematic uncertainties. The model for the evolution in $w(a)$ is given in \secref~\ref{sec:cosmomodel}. Gray-shaded contours show the constraints when the transition scale factor of $w(a)$ is fixed to $\atr=0.5$, while for the gold-shaded contours it is marginalized over the range $0.5<\atr<0.95$. The model corresponding to a cosmological constant is shown by a cross.
    Right: Joint constraints on $w_0$ and $\Omegam$ (blue) and $\wet$ and $\Omegam$ (purple) from the combination of data, for models with $\atr$ free (corresponding to the gold contours in the left panel). The dashed line, $w_0=\wet=-1$, corresponds to the cosmological constant model.
  }
  \label{fig:evolvingw}
\end{figure*}

 Allowing the parameter in \eqnref~\ref{eq:wdef} governing the evolution dark energy equation of state, $w_a$, to be free, we investigate the constraints available from the combination of \fgas{}, CMB, supernova and BAO data in two cases: fixing the transition scale factor at $\atr=0.5$ (i.e.\ the model is that of \citealt{Chevallier0009008} and \citealt{Linder0208512}) and marginalizing over the range $0.5<\atr<0.95$, as in \citet{Rapetti0409574} and \arsemf{}. The resulting constraints on $w_0$ and $w_a$ are shown in the left panel of \figref~\ref{fig:evolvingw} as gray and gold shaded contours, respectively. Curvature is allowed to vary, remaining tightly constrained and consistent with zero, in both cases. For completeness, \tabref~\ref{tab:death} shows results for models with both free and fixed curvature. In every case, the data are consistent with the \LCDM{} model ($w_0=-1$, $w_a=0$). The right panel of the figure shows the constraints on $w_0$ and $\wet=w_0+w_a$ versus \Omegam{} for models with curvature and $\atr$ free. Even for this general model, the combination of data provides a tight constraint on $\Omegam$, $0.294\pm0.017$.

\begin{table*}
  \begin{center}
    \caption{
      Marginalized best-fitting values and 68.3 per cent maximum-likelihood confidence intervals on cosmological parameters from our analysis, including systematic uncertainties. Parameters are defined in \secref~\ref{sec:cosmomodel}. The ``Comb$_\mathrm{WM}$'' combination of data refers to the union of our \fgas{} data set with CMB power spectra from WMAP \citep{Hinshaw1212.5226}, ACT \citep{Das1301.1037} and SPT \citep{Keisler1105.3182, Reichardt1111.0932, Story1210.7231}, the Union 2.1 compilation of type Ia supernovae \citep{Suzuki1105.3470}, and baryon acoustic oscillation measurements at $z=0.106$ \citep{Beutler1106.3366}, $z=0.35$ \citep{Padmanabhan1202.0090} and $z=0.57$ \citep{Anderson1303.4666}. ``Comb$_{Pl}$'' is identical, with the exception that 1-year \Planck{} data (plus WMAP polarization; \citealt{Planck1303.5075}) are used in place of the complete 9-year WMAP data. The \fgas{}-only constraints incorporate standard priors on $h$ and $\Omegab h^2$ (\tabref~\ref{tab:paramdef}; \citealt{Riess1103.2976, Cooke1308.3240}).
    }
    \label{tab:death}
    \vspace{1ex}
    \begin{tabular}{ccccccccc}
      \hline
      Model & Data & \Omegam{} & \Omegade{} & \phmin$\Omega_k$ & $w_0$ & $w_a$ & \wet{} & $\atr$ \\
      \hline\vspace{-1.5ex}\\
      \LCDM{} & \fgas{} & $0.29\pm0.04$ & $0.65^{+0.17}_{-0.22}$ & $0.08^{+0.19}_{-0.18}$ & $-1$ &  \phmin0 & &\vspace{1ex}\\
      & Comb$_\mathrm{WM}$ & $0.296\pm0.011$ & $0.706\pm0.012$ & $-0.003\pm0.004$ & $-1$ &  \phmin0 & &\vspace{1ex}\\
      & Comb$_{Pl}$ & $0.306\pm0.010$ & $0.695\pm0.010$ & $-0.001\pm0.003$ & $-1$ & \phmin0 & &\vspace{1ex}\\
      constant-$w$ & \fgas{} & $0.29\pm0.04$ &  & \phmin0 & $-0.98\pm0.26$ &  \phmin0 & &\vspace{1ex}\\
      & Comb$_\mathrm{WM}$  & $0.296\pm0.013$ &  & \phmin0 & $-1.02\pm0.08$ &  \phmin0 & &\vspace{1ex}\\
      & Comb$_{Pl}$  & $0.295\pm0.013$ &  & \phmin0 & $-1.08\pm0.07$ & \phmin0 & &\vspace{1ex}\\
      & Comb$_\mathrm{WM}$ & $0.291\pm0.014$ & $0.712\pm0.016$ & $-0.004\pm0.005$ & $-1.04\pm0.08$ &  \phmin0 & &\vspace{1ex}\\
      & Comb$_{Pl}$ & $0.292\pm0.014$ & $0.711\pm0.015$ & $-0.003\pm0.004$ & $-1.11\pm0.08$ & \phmin0 &  &\vspace{1ex}\\
      evolving-$w$ & \fgas{} & $0.28\pm0.04$ &  & \phmin0 & $-1.7\pm1.0$ & \phmin$2.2^{+2.4}_{-2.7}$ & \phmin$0.5^{+1.6}_{-1.8}$ & 0.5\vspace{1ex}\\
      & Comb$_\mathrm{WM}$ & $0.293\pm0.016$ & & \phmin0 & $-1.08\pm0.17$ & \phmin$0.4^{+0.5}_{-0.7}$ & $-0.8\pm0.4$ & 0.5\vspace{1ex}\\
      & Comb$_{Pl}$ & $0.298\pm0.015$ & & \phmin0 & $-1.03\pm0.18$ & $-0.1^{+0.6}_{-0.7}$ & $-1.1^{+0.4}_{-0.6}$ & 0.5\vspace{1ex}\\
      & Comb$_\mathrm{WM}$ & $0.295\pm0.016$ & $0.710\pm0.017$ & $-0.006\pm0.006$ & $-0.96\pm0.23$ & \phmin$0.0^{+0.9}_{-1.5}$ & $-1.1^{+0.7}_{-1.2}$ & 0.5\vspace{1ex}\\
      & Comb$_{Pl}$ & $0.304\pm0.017$ & $0.704\pm0.016$ & $-0.01\pm0.05$ & $-0.80\pm0.26$ & $-1.3^{+1.0}_{-1.5}$ & $-2.2^{+0.9}_{-1.2}$ & 0.5\vspace{1ex}\\
      & Comb$_\mathrm{WM}$ & $0.291\pm0.017$ &  & \phmin0 & $-1.11\pm0.26$ & \phmin$0.4^{+0.4}_{-0.7}$ & $-0.9\pm0.3$ & ---\vspace{1ex}\\
      & Comb$_{Pl}$ & $0.297\pm0.016$ &  & \phmin0 & $-1.02\pm0.25$ & \phmin$0.0^{+0.5}_{-0.6}$ & $-1.1^{+0.2}_{-0.3}$ & ---\vspace{1ex}\\
      & Comb$_\mathrm{WM}$ & $0.294\pm0.017$ & $0.709\pm0.017$ & $-0.006\pm0.006$ & $-0.99\pm0.34$ & \phmin$0.1^{+0.8}_{-1.0}$ & $-1.0^{+0.5}_{-0.7}$ & ---\vspace{1ex}\\
      & Comb$_{Pl}$ & $0.304\pm0.018$ & $0.702\pm0.017$ & $-0.008\pm0.005$ & $-0.75\pm0.34$ & $-0.9^{+0.9}_{-1.0}$ & $-1.5^{+0.5}_{-0.8}$ & ---\vspace{1ex}\\
   \hline
    \end{tabular}
  \end{center}
\end{table*}

\subsection{Impact of the \fgas{} Data}

As a simple measure of the influence of the \fgas{} data on our combined constraints, we compare the areas of the plotted 95.4 per cent confidence regions from the full combination of data to those obtained from combining only CMB, supernova and BAO data (i.e.\ excluding \fgas{}). For the \LCDM{} and flat, constant-$w$ models [respectively the $(\Omegam,\Omegal)$ and $(\Omegam,w)$ confidence regions] we find 11 per cent reductions in uncertainty when including the \fgas{} data in the combination. For the evolving-$w$ models (with free curvature), the allowed areas in the $(w_0,w_a)$ plane shrink by 34 per cent ($\atr=0.5$) and 29 per cent ($\atr$ free).

\section{Comparison to Previous Work} \label{sec:changes}

Having reported results from the present work, we now review the differences between our analysis and that of \arsemf{}, and their consequences for the cosmological constraints. 

\begin{figure*}
  \centering
  \includegraphics[scale=1]{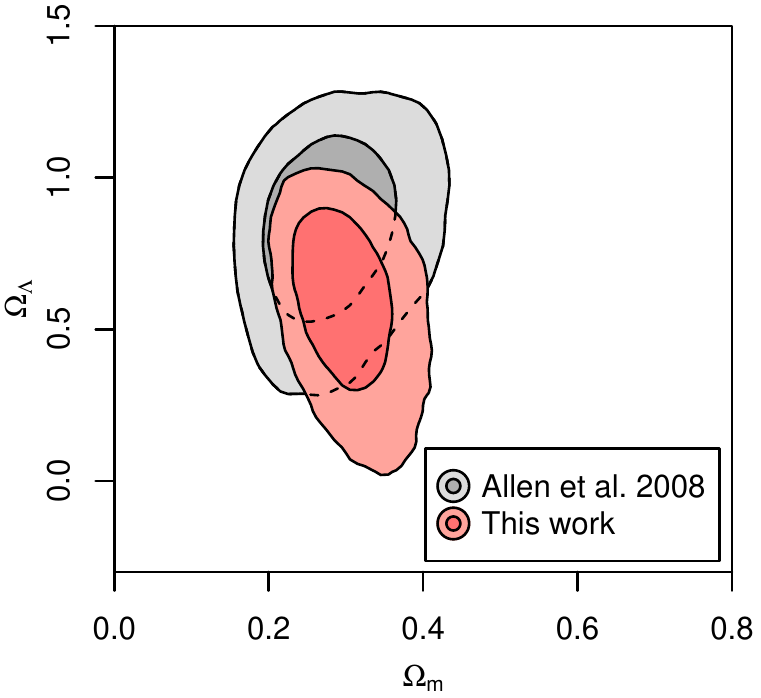}
  \hspace{1cm}
  \includegraphics[scale=1]{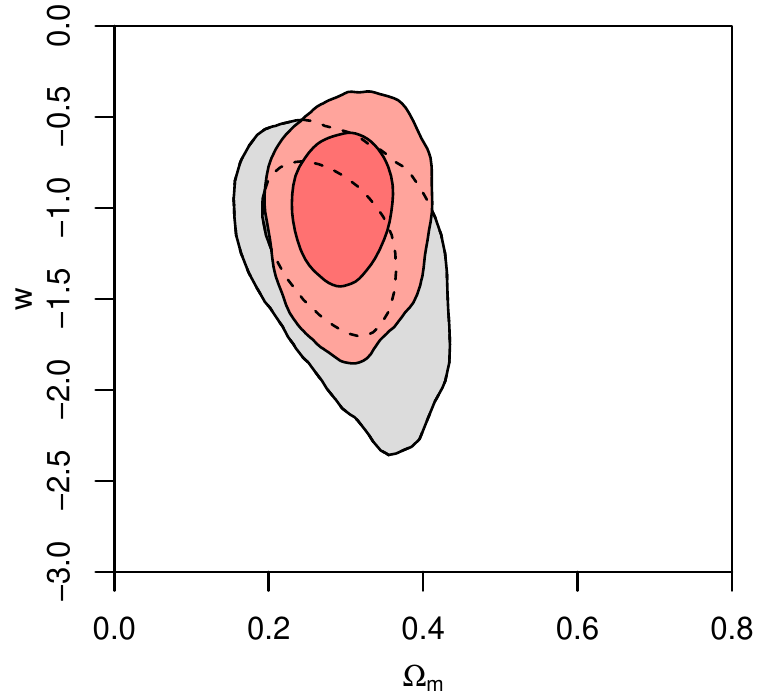}
  \caption{
    Cluster \fgas{} constraints on non-flat \LCDM{} (left) and flat constant-$w$ (right) models from the present analysis (red) and \arsemf{} (gray). Both results reflect contemporaneous priors on $h$ and $\Omegab h^2$ and allowances for systematic uncertainties. Dark and light shaded regions show the marginalized 68.3 and 95.4 per cent confidence regions.
  }
  \label{fig:a08}
\end{figure*}
 
\begin{enumerate}
\item The amount of Chandra data used has doubled: $3.1\Ms$ vs $1.6\Ms$, after cleaning.
\item The selection of relaxed clusters is now algorithmic rather than subjective. Although the present data set overlaps the \arsemf{} sample significantly and is almost the same size, roughly one quarter of our clusters were not represented in \arsemf{}. The turnover is particularly significant at redshifts 0.6--1.0, where MACS\,J0744, MS\,1137 and CL\,J1226 have been replaced by clusters discovered in the SPT survey.
\item In the present analysis, we use \fgas{} measured in a spherical shell at radii $0.8<r/r_{2500}<1.2$. This choice results in somewhat larger measurement uncertainties than we would obtain for the larger volume $r<r_{2500}$. However, it has the advantage of making the theoretical prior for the gas depletion significantly more robust to the particular implementation of gas physics in simulations. Our prior on the normalization of the depletion has a width of 20 per cent, compared to $\sim 40$ per cent previously. A related consequence of the use of this shell (excluding the cluster core) is that we can directly use simulated results for the {\it gas} depletion, rather than combining a prior on the {\it baryonic} depletion with measurements of the mass in stars relative to hot gas, without incurring additional systematic uncertainty.
\item \arsemf{} marginalized over priors for both instrument calibration (10 per cent Gaussian) and bias in mass measurements due to non-thermal pressure (10 per cent width uniform prior). In this work, we take advantage of recent improvements in weak gravitational lensing data and analysis methods (\wtg{}) to directly constrain the combination of these effects (see also Applegate et~al., in preparation).
\item In this work, our spectral analysis of each cluster (\secref{}s~\ref{sec:projct} and \ref{sec:nfwfit}) and subsequent cosmological analysis fully account for covariance between observables which are ultimately measured from the same photons (e.g.\ temperature and gas density, gas mass and total mass). This follows from the fact that we fit a single model for the mass, temperature and gas density profiles to the spectral data for each cluster. In contrast, most previous work (e.g.\ \citealt{LaRoque0604039, Vikhlinin0507092, Ettori0904.2740}) has involved fitting temperature profiles to X-ray spectra, deriving gas density profiles from X-ray surface brightness profiles, and then combining these (as if they were independent measurements) to constrain the mass profiles. Accounting for the measurement correlations slightly tightens the individual \fgas{} error bars and makes them more robust.
\item We use updated priors on $h$ and $\Omegab h^2$ \citep{Riess1103.2976, Cooke1308.3240}.
\end{enumerate}

Items 3 and 4 directly impact the precision and accuracy of \Omegam{} measurements from \fgas{} data. The reduction in uncertainty that results from working outside cluster centers and incorporating direct weak lensing mass calibration shrinks the width of our \Omegam{} constraint by $>30$ per cent relative to \arsemf{}. The priors on $h$ and $\Omegab h^2$ are sub-dominant in determining our final results. \figref~\ref{fig:a08} compares our new constraints with those of \arsemf{} for non-flat \LCDM{} and flat constant-$w$ models, highlighting in particular the improved constraint on \Omegam{}.

In contrast, our dark energy constraints are not markedly improved over those of \arsemf{}. There are two principal reasons for this. First, our cluster sample has not grown at high redshifts; the strict requirements for relaxation introduced in \morphpaper{} result in almost as many clusters at $z \gtsim 0.5$ being removed from the sample as new clusters have been added. A second factor is the presence of intrinsic scatter at the $\sim7.5$ per cent level, which we have detected here for the first time. Although this scatter is quite small, it implies that significant improvements will require the addition of new relaxed clusters to the data set, especially at redshifts where the current data are sparse (see also \secref~\ref{sec:future}).

Perhaps the most important consequence of the changes described above is that they greatly lower the systematic floor for the \fgas{} technique. Here we particularly emphasize the use of an optimized measurement shell as opposed to a sphere; the availability of X-ray/lensing mass calibration (\wtg; Applegate et~al., in preparation); the sample selection, codified in the morphological analysis of \morphpaper{}; and the blind analysis of both the X-ray and lensing data (\secref{}s~\ref{sec:projct} and \ref{sec:nfwfit}; \wtg). The latter aspects minimize the possibility of unconscious observer bias, providing an extra level of robustness to our results. Characterizing the intrinsic scatter is another critical step, both for \fgas{} cosmology and for the use of ICM observables as proxies for total mass. Together, these developments raise the prospect of substantial improvements in constraining power as more data are acquired, as we discuss in the next section.

\section{Prospects for Further Improvement} \label{sec:future}

\citet*{Rapetti0710.0440} studied the improvements in \fgas{} cosmology achievable in the context of the then-planned Constellation-X and XEUS observatories. However, that work underestimated the rate of progress in mitigating systematic uncertainties; in particular, the pessimistic scenarios considered by those authors can now be excluded. Additionally, our measurement of the intrinsic scatter in \fgas{} measurements impacts the observational strategy for future \fgas{} work. In a white paper based on a preliminary version of the work presented here (\citealt{Allen1307.8152}; see also \tabref~\ref{tab:fom}), we have revisited the subject of what improved constraints might be possible over the next 5--10 years using additional \Chandra{} and XMM-{\it Newton} observations, and on a longer timescale using a next generation, flagship X-ray observatory (hereafter NXO) coupled with Large Synoptic Survey Telescope (LSST)-like gravitational lensing data \citep{LSSTDESC1211.0310}. Currently proposed mission concepts include SMART-X\footnote{\url{http://hea-www.cfa.harvard.edu/SMARTX/}} and ATHENA+ \citep{Nandra1306.2307}. To be concrete, we consider the potential of an observatory with comparable spatial resolution to \Chandra{}, but $\sim30$ times the collecting area (akin to SMART-X).

For both possibilities, \Chandra{} and NXO, we simulate representative \fgas{} data sets that could be constructed from 10\,Ms of new observations, targeting clusters with redshifts drawn fairly from the expected distribution of systems with temperatures $>5\keV$ at redshifts $0.3<z<1.75$. See \citet{Allen1307.8152} for full details of the simulation procedure. With a 10\,Ms investment of \Chandra{} observing time over the next 5--10 years, 50 or more new clusters could be observed with exposures sufficient to measure \fgas{} to $\sim 15$ per cent precision, providing a final data set of nearly $100$ clusters, including current data. With 10\,Ms of observing time, an NXO with the capabilities described above could measure \fgas{} to 7.5 per cent precision for more than 400 clusters selected in the same way.

In order to keep the interpretation of projected cosmological results simple, we consider only two sets of priors and systematic allowances, corresponding to pessimistic and optimistic scenarios, where the pessimistic case generally assumes no improvement compared to the present. For simplicity, we have implemented the lensing/X-ray mass calibration as a redshift-independent prior applied to the simulated \fgas{} data, rather than simulating future weak lensing data sets. These priors are summarized in \tabref~\ref{tab:sys}.\footnote{The depletion prior (specifically on $\Upsilon_0$) that we use in this work is more conservative than the prior used by \citet{Allen1307.8152} for current data and the future-pessimistic simulations. As a result, we report slightly different figures of merit here compared to that work. Note also that the priors chosen for $h$ are compatible with recent calculations of the cosmic variance of the local Hubble constant based on large-scale cosmological simulations \citep{Wojtak1312.0276}.}

\begin{table}
  \begin{center}
    \caption{Systematic allowances on parameters used in projecting future constraints from \fgas{} data, expressed as fractions of their fiducial values.}
    \label{tab:sys}
    \begin{tabular}{ c c c c c c c }
      \hline
      Parameter            & & Current             & \multicolumn{2}{c}{Future}   &  Form    \\
      & & & pessimistic & optimistic & \\
      \hline
      $\Omega_{\rm b}h^2$ & &  $\pm0.02$  & $\pm0.02$&$\pm0.01$   & Gaussian \\
      $h$                 & &  $\pm0.03$  & $\pm0.03$&$\pm0.01$   & Gaussian\smallskip\\
      $K$                 & &  $\pm0.10$  & $\pm0.05$&$\pm0.02$   & Gaussian \\
      $\Upsilon_0$        & &  $\pm0.10$  & $\pm0.10$ &$\pm0.02$   & Uniform  \\
      $\Upsilon_1$ & &  $\pm0.05$  & $\pm0.05$&$\pm0.02$   & Uniform  \\
      \hline                      
    \end{tabular}
  \end{center}
\end{table}

In the pessimistic case, for both the 10\,Ms Chandra and NXO data sets, we incorporate intrinsic scatter at the current 7.5 per cent level. For the optimistic case with NXO, we consider the possibility that measurements of bulk and turbulent gas velocities with high resolution X-ray spectrometers will allow us to refine the selection of relaxed clusters further, reducing the intrinsic scatter to 5 per cent.

\begin{table*}
  \begin{center}
   \caption{Projected figures of merit for the \fgas{} experiment from simulations appropriate for current data plus 10\,Ms of new Chandra observations (93 clusters total) and 400 clusters observed with a next generation X-ray observatory (NXO) with 30 times the collecting area of Chandra. Our figure of merit is defined as the inverse of the area enclosed by the 95.4 per cent confidence contour for the associated pair of parameters, normalized by the constraints provided by current data. The fractional precision of the \Omegam{} constraint is also shown.}
    \label{tab:fom}
    \vspace{1ex}
    \begin{tabular}{ccc@{\hspace{2em}}cc@{\hspace{2em}}cc}
      \hline
     Model & Parameters & Priors & \multicolumn{2}{c}{\Chandra{} 10\,Ms~~~~~} & \multicolumn{2}{c}{NXO~~~~~~} \\
      &&& FoM$_{\rm c}$ & $\Delta\Omegam/\Omegam$ & FoM$_{\rm c}$ & $\Delta\Omegam/\Omegam$\\
      \hline
      non-flat \LCDM{} & \Omegam{} -- \Omegal{} & pessimistic & 2.2 &  0.09 & 3.2 & 0.08\\
      &   & optimistic & 6.0 & 0.05 & 14.8 & 0.03\smallskip\\
      flat $w$CDM & \Omegade{} -- $w$ & pessimistic & 2.3 & 0.09 & 3.6 & 0.08\\
      &   & optimistic & 6.6 & 0.05 & 15.0 & 0.03\smallskip\\
      flat evolving-$w$ & $w_0$ -- $w_a$ & pessimistic & 2.0 & 0.13 & 5.5 & 0.11\\
      &   & optimistic & 3.0 & 0.09 & 16.9 & 0.06\\
      \hline                      
    \end{tabular}
  \end{center}
\end{table*}

\begin{figure*}
  \centering
  \includegraphics[scale=0.75]{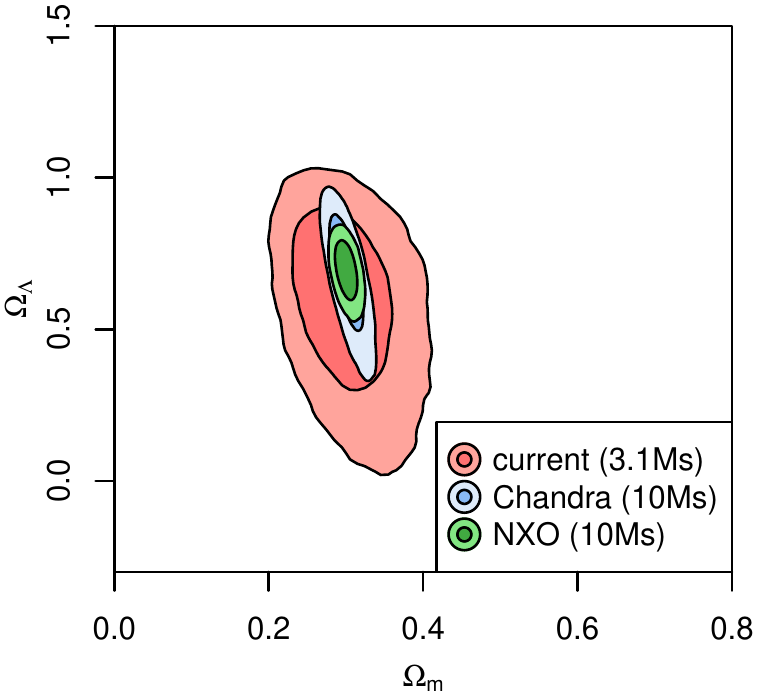}
  \includegraphics[scale=0.75]{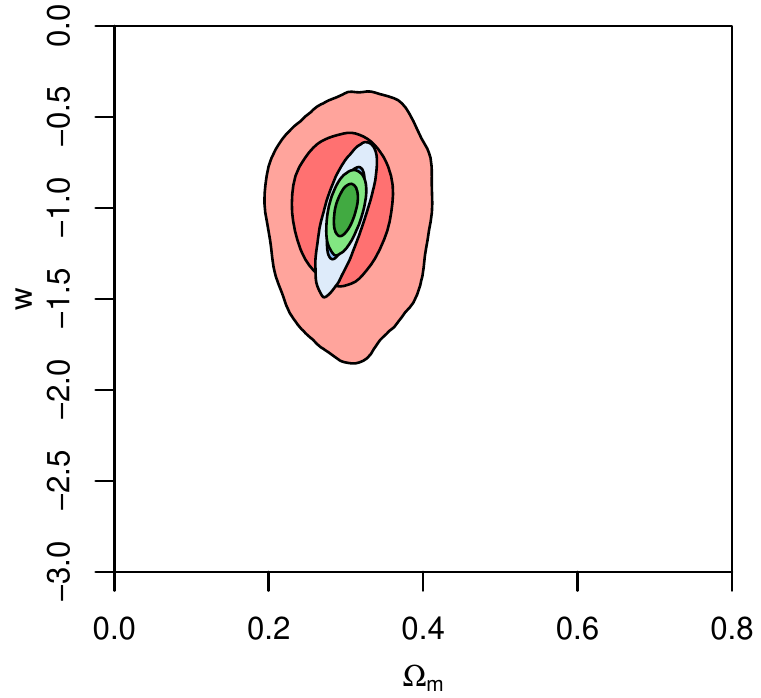}
  \includegraphics[scale=0.75]{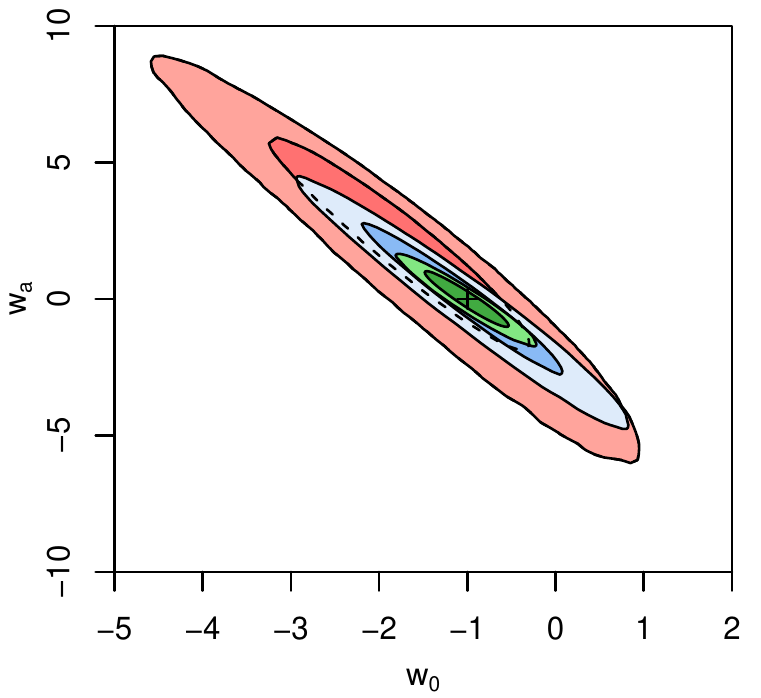}
  \caption{Joint 68.3 and 95.4 per cent confidence constraints from the \fgas{} test for non-flat \LCDM{} (left); flat, constant-$w$ (center); and flat, evolving-$w$ (right, with fixed $\atr=0.5$) models. Red shading shows the constraints  from current \fgas{} data.  Blue shading shows the predicted, improved constraints when adding 10\,Ms of \Chandra{} to provide \fgas{} measurements for 53 more clusters to a precision of $\sim 15$ per cent. Green contours show predicted constraints from combining current data with a future data set of 400 clusters with \fgas{} measured to 7.5 per cent precision using a new, next generation X-ray observatory with 30 times the collecting of \Chandra{}. Optimistic priors are assumed for the projections shown (see \tabref~\ref{tab:sys}).  The cross in the right panel marks the cosmological constant model ($w_0=-1$, $w_a=0$).}
  \label{fig:simres}
\end{figure*}

 We investigate three cosmological models, each of which can be fully constrained by \fgas{} data, plus priors on $h$ and $\Omegab h^2$: \LCDM{}, flat constant-$w$, and flat evolving-$w$ (with $\atr=0.5$). To quantify the improvements in cosmological constraining power for these models, we define our figure of merit as the inverse of the area enclosed by the 95.4 per cent confidence contour for the associated pair of parameters [($\Omegam,\Omegal$), ($\Omegade,w$) or ($w_0,w_a$), respectively], normalized to the constraints provided by current data.\footnote{The decision to normalize our figure of merit to the current \fgas{} constraints, and to use only \fgas{} data rather than incorporating simulated 2-year \Planck{} data (as in \citealt{Albrecht0609591}), makes our projections independent of external data, but arguably less easy to compare to other projections in the literature.}

\tabref~\ref{tab:fom} shows the predicted improvements in cosmological constraining power for both the simulated 10\,Ms \Chandra{} and NXO \fgas{} data sets. In addition to the figures of merit, we include the fractional uncertainty in the marginalized constraint on \Omegam, which is relatively insensitive to the choice of dark energy model ($\sim5$ per cent precision, optimistically). The corresponding two-dimensional confidence regions (blue/green for \Chandra{}/NXO, respectively) are compared with our current results (red) in \figref~\ref{fig:simres}. Note that only cluster \fgas{} data, in conjunction with priors on $\Omegab h^2$ and $h$, are used here. Constraints from the simulated 10\,Ms \Chandra{} data set are improved with respect to current data by factors of 2--7; for the NXO data set, in the optimistic case, improvement factors of 15--17 are found. The impact is greatest for the evolving-$w$ model, where the NXO figure of merit is a further factor of 5--6 better than that from the simulated 10\,Ms \Chandra{} data set. These tight constraints highlight the potential for X-ray cluster observations to provide competitive constraints on cosmic distances, complementary to those of other probes, going forward. However, realizing the full potential of new data will also require continuing improvements in hydrodynamic simulations, gravitational lensing measurements, and external constraints on the Hubble parameter and cosmic baryon density.

\section{Conclusions} \label{sec:cosmosummary}

We have presented cosmological constraints from X-ray gas mass fraction measurements of a sample of hot, massive, dynamically relaxed galaxy clusters. This study builds on the previous work of \arsemf{} in several respects. In addition to incorporating roughly twice as much \Chandra{} data, our selection of morphologically relaxed clusters has now been automated (\morphpaper{}). The present sample incorporates all sufficiently hot, dynamically relaxed systems with adequate exposures found in a comprehensive search of the \Chandra{} archive. Systematic uncertainties have been reduced by using measurements of \fgas{} in spherical shells that exclude cluster centers, where theoretical predictions are most uncertain, and by using gravitational lensing data to directly constrain systematic uncertainties associated with non-thermal pressure support and instrument calibration (Applegate et~al., in preparation). Throughout the target selection process, the X-ray analysis of individual clusters, and the lensing/X-ray calibration, the analysis team was blinded to all results that could influence the final cosmological interpretation.

The reductions in systematic uncertainty principally affect the constraint on \Omegam{} that follows from the normalization of the $\fgas(z)$ curve, a measurement that, importantly, is largely insensitive to assumptions about the nature of dark energy. Our constraints on dark energy are similar to those of \arsemf{} due to the similar size and redshift distribution of the cluster samples used. The results from this work are $\Omegam=0.29\pm0.04$, $\Omegal=0.65^{+0.17}_{-0.22}$ for non-flat \LCDM{} models, and $w=-0.98\pm0.26$ for flat, constant-$w$ models (with an identical constraint on $\Omegam$). Combining with CMB, supernova and BAO data, we find tighter constraints that remain consistent with the value $\Omegam\sim0.3$ preferred by the \fgas{} data, as well as with the cosmological constant model, even in models with free global curvature and evolving $w(z)$ (\tabref~\ref{tab:death}).

The high precision of our \Chandra{} data permit us to detect, for the first time, the intrinsic scatter in \fgas{} measurements for these highly relaxed clusters. The fractional intrinsic scatter, $0.074\pm0.023$ in the 0.8--1.2\,$r_{2500}$ measurement shell, corresponds to a systematic uncertainty of only $\sim5$ per cent in the cosmological distance to a given cluster. This small scatter (as well as the tight constraint on $\Omegam$, essentially independent of the dark energy model considered) explains why dark energy constraints from \fgas{} data remain competitive with those of, for example, type Ia supernovae \citep{Suzuki1105.3470}, despite the fact that typical supernova data sets are now an order of magnitude larger than our relaxed cluster sample. The measured scatter places a limit on the variation in non-thermal pressure in these relaxed clusters, which future, larger lensing and X-ray data sets may be able to constrain directly. In the near term, observations with the upcoming ASTRO-H mission \citep{Takahashi1010.4972} should provide critical insights into the degree of turbulent and bulk gas motions in nearby clusters. Farther ahead, a high-resolution X-ray microcalorimeter aboard a new flagship observatory should allow refined \fgas{} measurements for large samples of relaxed clusters and provide gas velocity information to potentially reduce the intrinsic scatter in \fgas{} measurements.

Significant improvement in dark energy constraints from the \fgas{} method will require the discovery of new relaxed clusters at redshifts $z>0.5$ from upcoming surveys, as well as a significant investment of time by flagship X-ray telescopes to observe the new targets (e.g., initially with \Chandra{}, for the brightest new sources, and later with a next-generation observatory). We project that factors $>15$ improvement in constraining power could be achieved over the next $\sim20$ years, given a sustained observing program. Realizing this potential will also require significant, but entirely plausible, reductions in systematic uncertainties through continued refinement of hydrodynamic simulations, and expanding the high-quality gravitational lensing data available for relaxed clusters.

Our data and likelihood code are available at\\ \url{http://www.slac.stanford.edu/~amantz/work/fgas14/}.

\section*{Acknowledgments}

The authors wish to thank Paul Nulsen for providing an advance copy of the {\sc clmass} code, Craig Gordon and Keith Arnaud for their work on {\sc xspec}, Nick Battaglia and Susana Planelles for sharing details of their cluster simulations, and Peter Behroozi, Andrey Kravtsov and Sam Skillman for helpful insights. Calculations for this work utilized the Coma, Orange and Bullet compute clusters at the SLAC National Accelerator Laboratory.

AM was supported by National Science Foundation grants AST-0838187 and AST-1140019. DA acknowledges funding from the German Federal Ministry of Economics and Technology (BMWi) under project 50 OR 1210. We acknowledge support from the U.S. Department of Energy under contract number DE-AC02-76SF00515; from the National Aeronautics and Space Administration (NASA) through Chandra Award Numbers GO8-9118X and TM1-12010X, issued by the Chandra X-ray Observatory Center, which is operated by the Smithsonian Astrophysical Observatory for and on behalf of NASA under contract NAS8-03060; as well as through program HST-AR-12654.01-A, provided by NASA through a grant from the Space Telescope Science Institute, which is operated by the Association of Universities for Research in Astronomy, Inc., under NASA contract NAS 5-26555. The Dark Cosmology Centre is funded by the Danish National Research Foundation.

\def \aap {A\&A} 
\def \aapr {A\&AR} 
\def \aaps {A\&AS} 
\def \statisci {Statis. Sci.} 
\def \physrep {Phys. Rep.} 
\def \pre {Phys.\ Rev.\ E} 
\def \sjos {Scand. J. Statis.} 
\def \jrssb {J. Roy. Statist. Soc. B} 
\def \pan {Phys. Atom. Nucl.} 
\def \epja {Eur. Phys. J. A} 
\def \epjc {Eur. Phys. J. C} 
\def \jcap {J. Cosmology Astropart. Phys.} 
\def \ijmpd {Int.\ J.\ Mod.\ Phys.\ D} 
\def \nar {New Astron. Rev.}

\def \araa {ARA\&A}
\def \aj {AJ}
\def \aar {A\&AR}
\def \apj {ApJ}
\def \apjl {ApJL}
\def \apjs {ApJS}
\def \asl {Adv. Sci. Lett.} 
\def \mnras {MNRAS}
\def \nat {Nat}
\def \pasj {PASJ}
\def \pasp {PASP}
\def \science {Sci}
\def \compcom {Comput.\ Phys.\ Commun.}
\def \gca {Geochim.\ Cosmochim.\ Acta}
\def \npa {Nucl.\ Phys.\ A}
\def \plb {Phys.\ Lett.\ B}
\def \prc {Phys.\ Rev.\ C}
\def \prd {Phys.\ Rev.\ D}
\def \prl {Phys.\ Rev.\ Lett.}

\appendix

\section{Practical Details of the {\sc nfwmass} Model} \label{sec:nfwmass}

As described in \secref~\ref{sec:nfwfit}, the {\sc nfwmass} model which we fit to our spectral data makes no assumptions regarding cosmology. Consequently, model-dependent factors must be applied in order to convert the fitted parameter values to physically meaningful masses, gas densities, etc. We review those somewhat complex conversion factors in this appendix.

In the {\sc nfwmass} model, the cluster atmosphere is described as a set of concentric, spherical shells, with radii corresponding to the set of annuli from which spectra are being analyzed. The full set of parameters is $\{\xsrs$, $\xsmod{\Phi}_0$, $K_1$, $kT_1$, \ldots, $kT_N\}$. Here $K_1$ is the \sw{xspec} normalization of the innermost shell, which is related to the gas density profile normalization; $kT_i$ is the temperature of the $i$th shell in keV; and $N$ is the number of shells/annuli. $\xsrs$ and $\xsmod{\Phi}_0$ parametrize the gravitational potential of the \NFW{} model:
\begin{equation}
  \xsmod{\Phi}(x) = \xsmod{\Phi}_0 \left[ 1 - \frac{\ln(1+x)}{x} \right],
\end{equation}
where $x$ is radius in units of $\xsrs$.
The corresponding mass profile is
\begin{equation}
  \xsmod{M}(x) = \xsmod{\Phi}_0 \, \xsrs \left[  \ln(1+x) - \frac{x}{1+x} \right].
\end{equation}
Tildes over these symbols indicate that they are not in physically meaningful units, as the cosmology-dependent conversions have not yet been applied. In particular, the unit of angular radius for \xsrs{}, $p$, is up to the user, and appears explicitly in the final conversion factors.

The model gas density at the inner edge of the innermost shell is given by (still in unphysical units)
\begin{equation}
  \xsmod{n}_\mathrm{i}^{-2} = \frac{4 \pi}{K_1} \int_{\xsmod{r}_\mathrm{i}}^{\xsmod{r}_\mathrm{o}} dy \,y \sqrt{y^2 - \xsmod{r}_\mathrm{i}^2},
\end{equation}
where $\xsmod{r}_\mathrm{i}$ and $\xsmod{r}_\mathrm{o}$ are the inner and outer radii of the innermost shell. Given the set of shell temperatures and the mass profile, the corresponding model density at any radius can be calculated as described by \citet{Nulsen1008.2393}.

Given a redshift and a cosmological model specifying $d(z)$, $\xsmod{M}$ and $\xsmod{n}$ are related to mass and particle density by
\begin{eqnarray}
  \xsmod{M}   & = & \frac{1}{p\keV}\left[ \frac{G\mu\mproton }{\dA(z)} \right] M, \nonumber \\
  \xsmod{n}^2 & = & \frac{p^3 \cm^{5}}{4\pi\E{14}} \left[ (1+z)^2 \frac{\dA(z)^3}{\dL(z)^2} \right] \nelec\nH.
\end{eqnarray}
Here $G$ is Newton's constant, $\mu\mproton$ is the mean molecular mass of the ICM, $\dL$ and \dA{} are the luminosity and angular diameter distances to the cluster, and $\nelec$ and $\nH$ are the number densities of free electrons and protons. The mean molecular mass has a weak dependence on the cosmological model, ultimately through the cosmic baryon density. Neglecting the contribution of nuclei heavier than helium, we can write
\begin{equation}
  \mu = \frac{\nelec}{\ntot} \left( \frac{\nH}{\nelec} + \mHe \frac{\nHe}{\nelec} \right),
\end{equation}
where $\ntot$ is the total particle number density of the plasma. The various number densities are related by
\begin{eqnarray}
  \frac{\nelec}{\nH} &=& 1 + 2\frac{\nHe}{\nH}, \nonumber \\
  \frac{\ntot}{\nelec} &=& 1 + \frac{1 + \nHe/\nH}{\nelec/\nH}, \nonumber \\
  \frac{\nHe}{\nH} &=& \frac{\YHe \, \mH}{(1-\YHe)\mHe},
\end{eqnarray}
where $\YHe$ is the primordial mass fraction of helium, which is related to $\Omegab h^2$ by the theory of BBN (e.g.\ \citealt{Pisanti0705.0290}). Making use of the above results, the conversion between physical gas mass to that computed by integrating $\xsmod{n}(\xsmod{r})$ is
\begin{equation}
  \xsmod{M}_\mathrm{gas} = \sqrt{\frac{\mathrm{cm}^5}{4\pi\E{14}\,p^3}} ~ \left[ \frac{\sqrt{\nelec\nH}\,(1+z)}{\mu\mproton\,\ntot\,\dL(z)\,\dA(z)^{3/2}} \right] \Mgas.
\end{equation}
The corresponding relation for \fgas{} is thus
\begin{eqnarray}
  \xsmod{f}_\mathrm{gas} & = & \frac{\mathrm{keV}\cm^{5/2}}{4\pi\E{14}\,p} \left( \frac{1}{G(\mu\mproton)^2} \right) \left( \frac{\sqrt{\nelec\nH}}{\ntot} \right) \nonumber\\
  & & \times \left( \frac{1+z}{\dL(z)\,\dA(z)^{1/2}} \right) \fgas.
\end{eqnarray}

\section{Figures Using \Planck{} Data} \label{sec:planckres}

\begin{figure*}
  \centering
  \includegraphics[scale=1]{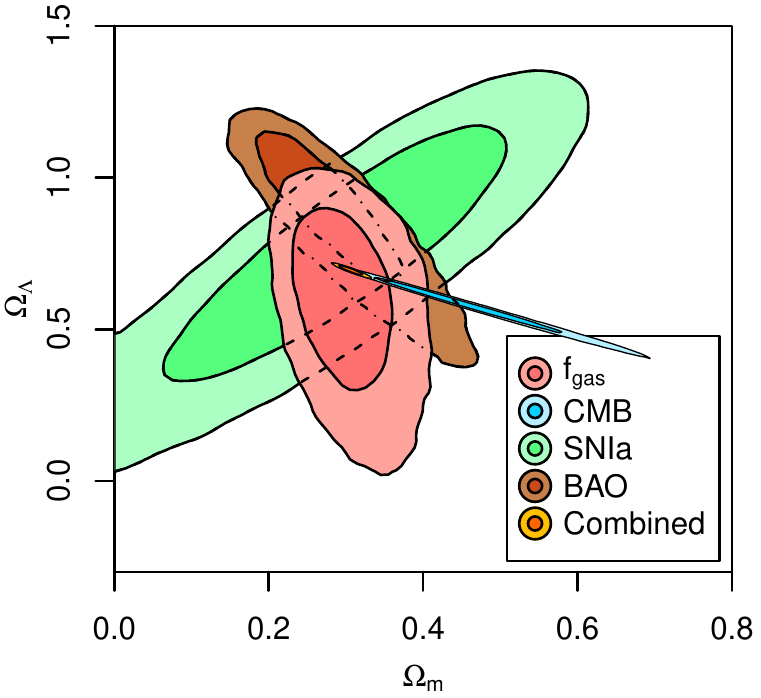}
  \hspace{1cm}
  \includegraphics[scale=1]{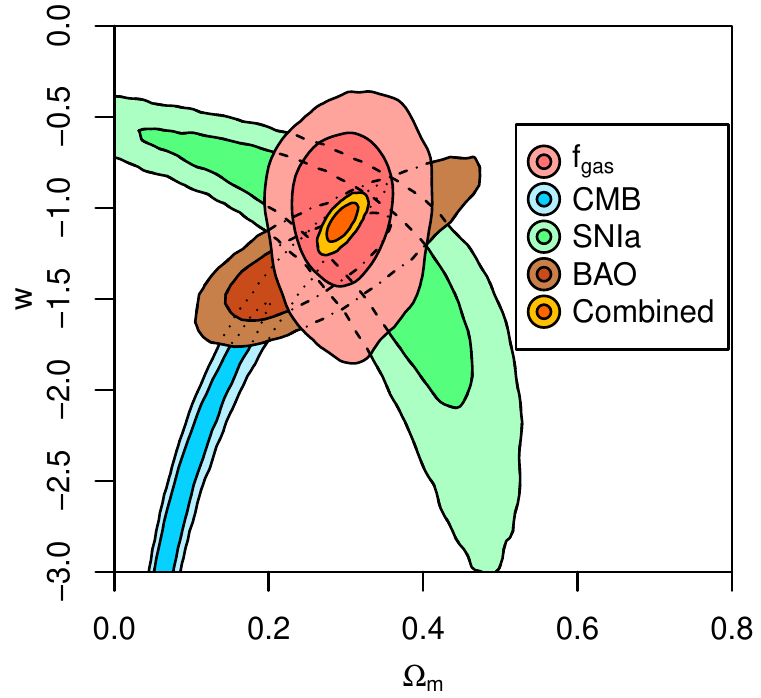}\\
  \includegraphics[scale=1]{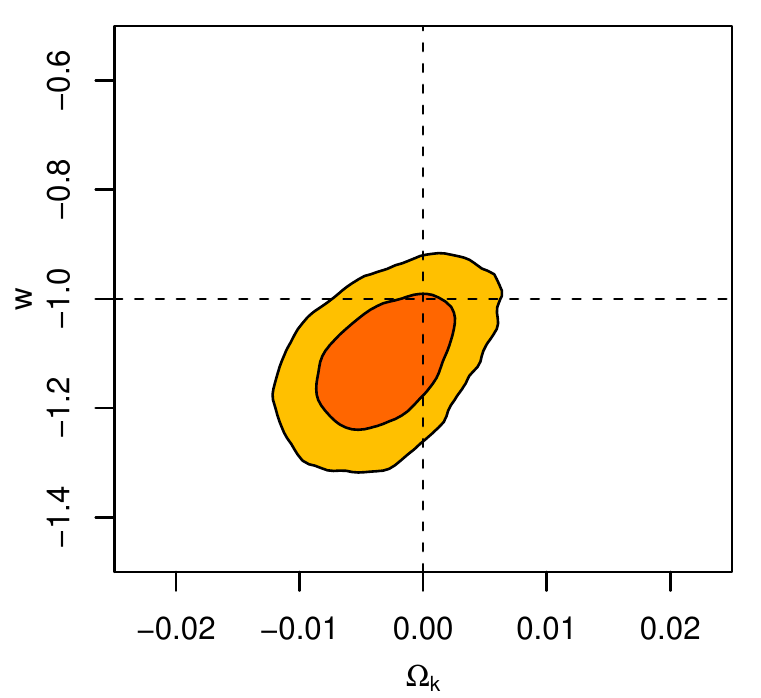}
  \hspace{1cm}
  \includegraphics[scale=1]{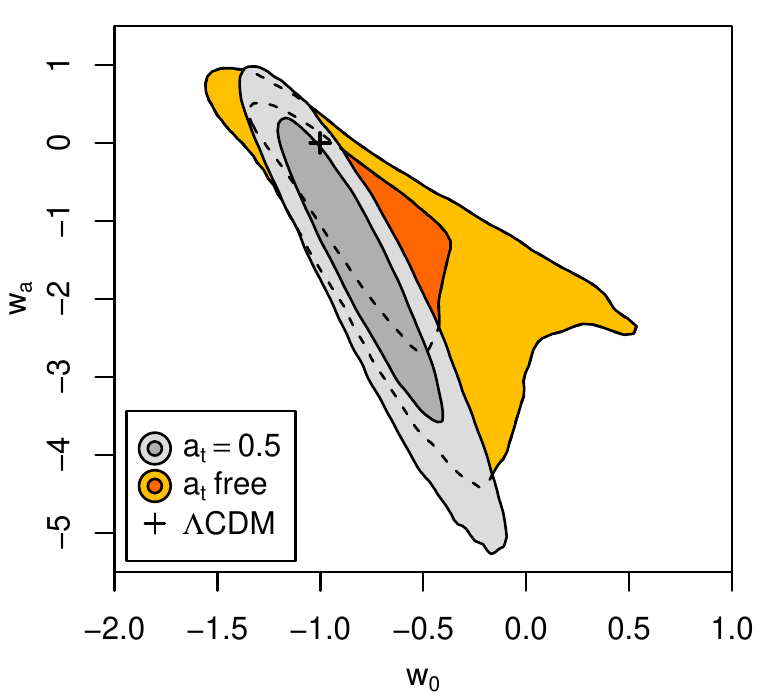}
  \caption{
    Constraints on cosmological models from the full cluster \fgas{} data set (red, including standard priors on $h$ and $\Omegab h^2$), CMB data from \Planck{}, ACT and SPT (blue; \citealt{Keisler1105.3182, Reichardt1111.0932, Story1210.7231, Das1301.1037,Planck1303.5076}), type Ia supernovae (green; \citealt{Suzuki1105.3470}), baryon acoustic oscillations (brown, also including priors on $h$ and $\Omegab h^2$; \citealt{Beutler1106.3366, Padmanabhan1202.0090, Anderson1303.4666}), and the combination of all four (gold). The priors on $h$ and $\Omegab h^2$ are not included in the combined constraints. These figures are identical to those in \secref~\ref{sec:cosmores} apart from the substitution of \Planck{} 1-year data (plus WMAP polarization) for WMAP 9-year data. Left to right and top to bottom: \LCDM{} models; flat, constant-$w$ models; non-flat, constant-$w$ models; non-flat, evolving-$w$ models.
  }
  \label{fig:planckfigs}
\end{figure*}

\figref{}~\ref{fig:planckfigs} shows the results equivalent to \figref{}s~\ref{fig:lcdm}--\ref{fig:evolvingw}a, with the substitution of \Planck{} 1-year data (plus WMAP polarization; \citealt{Planck1303.5076}) for WMAP 9-year data \citep{Hinshaw1212.5226}.

\bsp
\label{lastpage}
\end{document}

%% file: fgas_cluster_table.tex
\begin{table*}
  \centering
  \caption{%
    Galaxy clusters in our data set. Column [1] name; 
    [2], [3] J2000 coordinates of our adopted cluster center;
    [4] Galactic equivalent hydrogen column density;
    [5] clean \Chandra{} exposure time;
    [6] central exclusion radius (arcsec);
    [7] excluded position angles, if any (degrees);
    [8] whether a Galactic foreground component is included in the spectral modeling of each cluster;
    [9] whether each cluster is in the \wtg{} weak lensing sample, which forms the basis of our absolute mass calibration.
    Column densities are from the Leiden/Argentine/Bonn survey \citep{Kalberla0504140}, except where errors are shown, in which case they were fitted to the X-ray data. Note that the column density for Abell~478 is treated specially, as described in the main text. Redshifts appear in \tabref~\ref{tab:fgas}.
  }
  \begin{tabular}{lcccrrccc}
    \hline
    \multicolumn{1}{l}{Cluster}  & RA & Dec & $\NH$ & \multicolumn{1}{c}{exp.} & \multicolumn{2}{c}{Exclusion} & fg & \wtg{} \\
    & & & ($10^{20}\cm^{-2}$) & \multicolumn{1}{c}{(ks)} & radius & angle \\
    \hline
    Abell~2029           &  15:10:55.9  &  +05:44:41.2    &  3.26          &  118.6  &  39.4  &  ~          & $\surd$ &         \\
    Abell~478            &  04:13:25.2  &  +10:27:58.6    &  $16.8\pm2.0$  &  131.2  &  43.3  &  ~          &         &         \\
    RX~J1524.2$-$3154    &  15:24:12.8  &  $-$31:54:24.3  &  8.53          &  40.9   &  31.5  &  ~          & $\surd$ &         \\
    PKS~0745$-$191       &  07:47:31.7  &  $-$19:17:45.0  &  $54.8\pm0.3$  &  152.9  &  31.5  &  ~          & $\surd$ &         \\
    Abell~2204           &  16:32:47.1  &  +05:34:31.4    &  5.67          &  89.4   &  23.6  &  ~          & $\surd$ & $\surd$ \\
    RX~J0439.0+0520      &  04:39:02.3  &  +05:20:43.6    &  8.92          &  34.7   &  15.7  &  ~          &         &         \\
    Zwicky~2701          &  09:52:49.2  &  +51:53:05.3    &  0.75          &  111.3  &  17.7  &  ~          &         &         \\
    RX~J1504.1$-$0248    &  15:04:07.6  &  $-$02:48:16.7  &  5.97          &  39.9   &  13.8  &  ~          & $\surd$ &         \\
    Zwicky~2089          &  09:00:36.9  &  +20:53:40.4    &  2.86          &  47.0   &  13.8  &  ~          &         & $\surd$ \\
    RX~J2129.6+0005      &  21:29:39.9  &  +00:05:18.3    &  3.63          &  36.8   &  27.6  &  ~          & $\surd$ &         \\
    RX~J1459.4$-$1811    &  14:59:28.7  &  $-$18:10:45.0  &  7.38          &  39.6   &  39.4  &  ~          & $\surd$ &         \\
    Abell~1835           &  14:01:02.0  &  +02:52:39.0    &  2.04          &  205.3  &  25.6  &  ~          & $\surd$ & $\surd$ \\
    Abell~3444           &  10:23:50.2  &  $-$27:15:25.1  &  5.57          &  35.7   &  23.6  &  ~          & $\surd$ &         \\
    MS~2137.3$-$2353     &  21:40:15.2  &  $-$23:39:40.0  &  3.76          &  63.2   &  10.8  &  ~          & $\surd$ & $\surd$ \\
    MACS~J0242.5$-$2132  &  02:42:35.9  &  $-$21:32:25.9  &  2.72          &  7.7    &  11.8  &  ~          &         &         \\
    MACS~J1427.6$-$2521  &  14:27:39.5  &  $-$25:21:03.4  &  5.88          &  41.2   &  9.8   &  ~          & $\surd$ &         \\
    MACS~J2229.7$-$2755  &  22:29:45.2  &  $-$27:55:36.0  &  1.35          &  25.1   &  11.8  &  ~          &         &         \\
    MACS~J0947.2+7623    &  09:47:12.9  &  +76:23:13.8    &  2.28          &  48.3   &  10.8  &  ~          &         &         \\
    MACS~J1931.8$-$2634  &  19:31:49.6  &  $-$26:34:32.7  &  8.31          &  103.8  &  13.8  &  ~          & $\surd$ &         \\
    MACS~J1115.8+0129    &  11:15:51.9  &  +01:29:54.3    &  4.34          &  45.3   &  11.8  &  ~          &         & $\surd$ \\
    MACS~J1532.8+3021    &  15:32:53.8  &  +30:20:58.9    &  2.30          &  102.2  &  9.8   &  ~          &         & $\surd$ \\
    MACS~J0150.3$-$1005  &  01:50:21.3  &  $-$10:05:29.9  &  2.64          &  26.1   &  11.8  &  ~          & $\surd$ &         \\
    MACS~J0011.7$-$1523  &  00:11:42.9  &  $-$15:23:22.0  &  1.85          &  50.7   &  9.8   &  ~          &         &         \\
    MACS~J1720.2+3536    &  17:20:16.8  &  +35:36:27.0    &  3.46          &  53.2   &  9.8   &  235--355   &         & $\surd$ \\
    MACS~J0429.6$-$0253  &  04:29:36.1  &  $-$02:53:07.5  &  4.33          &  19.3   &  9.8   &  ~          & $\surd$ & $\surd$ \\
    MACS~J0159.8$-$0849  &  01:59:49.3  &  $-$08:50:00.1  &  2.06          &  62.3   &  19.7  &  ~          &         &         \\
    MACS~J2046.0$-$3430  &  20:46:00.6  &  $-$34:30:17.5  &  4.59          &  43.3   &  7.9   &  ~          & $\surd$ &         \\
    IRAS~09104+4109      &  09:13:45.5  &  +40:56:28.4    &  1.42          &  69.0   &  8.9   &  ~          &         &         \\
    MACS~J1359.1$-$1929  &  13:59:10.2  &  $-$19:29:23.4  &  5.99          &  54.7   &  7.9   &  ~          & $\surd$ &         \\
    RX~J1347.5$-$1145    &  13:47:30.6  &  $-$11:45:10.0  &  4.60          &  67.3   &  8.9   &  180--280   & $\surd$ & $\surd$ \\
    3C~295               &  14:11:20.5  &  +52:12:10.0    &  1.34          &  90.4   &  8.9   &  ~          &         &         \\
    MACS~J1621.3+3810    &  16:21:24.8  &  +38:10:09.0    &  1.13          &  134.0  &  9.8   &  ~          &         & $\surd$ \\
    MACS~J1427.2+4407    &  14:27:16.2  &  +44:07:31.0    &  1.19          &  51.0   &  7.9   &  250--370   &         & $\surd$ \\
    MACS~J1423.8+2404    &  14:23:47.9  &  +24:04:42.3    &  2.24          &  123.0  &  6.9   &  ~          & $\surd$ & $\surd$ \\
    SPT~J2331$-$5051     &  23:31:51.2  &  $-$50:51:54.0  &  1.12          &  31.8   &  3.9   &  ~          & $\surd$ &         \\
    SPT~J2344$-$4242     &  23:44:43.9  &  $-$42:43:13.0  &  1.52          &  10.7   &  6.9   &  ~          & $\surd$ &         \\
    SPT~J0000$-$5748     &  00:00:60.0  &  $-$57:48:33.6  &  1.37          &  28.4   &  3.9   &  ~          &         &         \\
    SPT~J2043$-$5035     &  20:43:17.6  &  $-$50:35:32.0  &  2.38          &  73.8   &  5.9   &  ~          & $\surd$ &         \\
    CL~J1415.2+3612      &  14:15:11.0  &  +36:12:02.6    &  1.05          &  348.8  &  3.9   &  ~          &         &         \\
    3C~186               &  07:44:17.5  &  +37:53:17.0    &  5.11          &  213.8  &  3.4   &  270--350   & $\surd$ &         \\
  \end{tabular}
  \label{tab:targets}
\end{table*}

%% file: fgas_table.tex
\begin{table*}
  \centering
  \caption{
    Redshifts, radii, masses, and \fgas{} (in the 0.8--1.2\,$r_{2500}$ shell) from our analysis. The listed radius, mass and \fgas{} values are calculated for our reference \LCDM{} cosmology. Quoted error bars are at the 68.3 per cent confidence level and include statistical uncertainties only. In particular, these values do not account for the measured offset between X-ray and gravitational lensing masses, or its uncertainty (\secref~\ref{sec:miscmodel}). The \fgas{} values are, however, marginalized over the (statistical) uncertainty in $r_{2500}$. SPT cluster redshifts are from \citet{Reichardt1203.5775} and \citet{McDonald1208.2962}.
  }
  \label{tab:fgas}
  \begin{tabular}{lclcc}
    \hline
    Cluster & $z$ & $r_{2500}^\mathrm{ref}$ & $M_{2500}^\mathrm{ref}$ & $\fgas ^\mathrm{ref}$ \\
    & & (kpc) & ($10^{14}\Msun$) & (0.8--1.2\,$r_{2500}$) \\ 
    \hline 
    Abell~2029           &  0.078  &  $662^{+5}_{-5}$    &                $4.41\pm0.10$   &  $0.131\pm0.003$\vspace{0.5ex}\\
    Abell~478            &  0.088  &  $634^{+7}_{-11}$   &                $3.88\pm0.17$   &  $0.128\pm0.008$\vspace{0.5ex}\\
    PKS~0745$-$191       &  0.103  &  $691^{+8}_{-5}$    &                $5.17\pm0.15$   &  $0.119\pm0.004$\vspace{0.5ex}\\
    RX~J1524.2$-$3154    &  0.103  &  $484^{+8}_{-9}$    &                $1.76\pm0.09$   &  $0.125\pm0.009$\vspace{0.5ex}\\
    Abell~2204           &  0.152  &  $707^{+12}_{-14}$  &                $5.73\pm0.31$   &  $0.131\pm0.008$\vspace{0.5ex}\\
    RX~J0439.0+0520      &  0.208  &  $497^{+14}_{-16}$  &                $2.12\pm0.20$   &  $0.111\pm0.015$\vspace{0.5ex}\\
    Zwicky~2701          &  0.214  &  $487^{+9}_{-5}$    &                $2.03\pm0.09$   &  $0.109\pm0.006$\vspace{0.5ex}\\
    RX~J1504.1$-$0248    &  0.215  &  $705^{+16}_{-16}$  &                $6.15\pm0.42$   &  $0.108\pm0.007$\vspace{0.5ex}\\
    RX~J2129.6+0005      &  0.235  &  $562^{+12}_{-14}$  &                $3.17\pm0.24$   &  $0.140\pm0.016$\vspace{0.5ex}\\
    Zwicky~2089          &  0.235  &  $442^{+9}_{-10}$   &                $1.53\pm0.10$   &  $0.127\pm0.012$\vspace{0.5ex}\\
    RX~J1459.4$-$1811    &  0.236  &  $570^{+11}_{-21}$  &                $3.25\pm0.29$   &  $0.130\pm0.010$\vspace{0.5ex}\\
    Abell~1835           &  0.252  &  $671^{+9}_{-8}$    &                $5.51\pm0.22$   &  $0.120\pm0.007$\vspace{0.5ex}\\
    Abell~3444          &  0.253  &  $561^{+12}_{-12}$  &                $3.23\pm0.22$   &  $0.142\pm0.013$\vspace{0.5ex}\\
    MS~2137.3$-$2353     &  0.313  &  $477^{+10}_{-9}$   &                $2.12\pm0.13$   &  $0.137\pm0.011$\vspace{0.5ex}\\
    MACS~J0242.5$-$2132  &  0.314  &  $528^{+35}_{-25}$  &                $3.00\pm0.56$   &  $0.125\pm0.028$\vspace{0.5ex}\\
    MACS~J1427.6$-$2521  &  0.318  &  $450^{+17}_{-18}$  &                $1.78\pm0.21$   &  $0.131\pm0.024$\vspace{0.5ex}\\
    MACS~J2229.7$-$2755  &  0.324  &  $481^{+14}_{-14}$  &                $2.18\pm0.20$   &  $0.133\pm0.016$\vspace{0.5ex}\\
    MACS~J0947.2+7623    &  0.345  &  $603^{+20}_{-16}$  &                $4.53\pm0.42$   &  $0.104\pm0.011$\vspace{0.5ex}\\
    MACS~J1931.8$-$2634  &  0.352  &  $584^{+12}_{-15}$  &                $4.04\pm0.28$   &  $0.112\pm0.011$\vspace{0.5ex}\\
    MACS~J1115.8+0129    &  0.355  &  $559^{+18}_{-10}$  &                $3.65\pm0.28$   &  $0.145\pm0.017$\vspace{0.5ex}\\
    MACS~J0150.3$-$1005  &  0.363  &  $431^{+18}_{-14}$  &                $1.67\pm0.19$   &  $0.152\pm0.021$\vspace{0.5ex}\\
    MACS~J1532.8+3021    &  0.363  &  $574^{+12}_{-11}$  &                $3.90\pm0.23$   &  $0.108\pm0.006$\vspace{0.5ex}\\
    MACS~J0011.7$-$1523  &  0.378  &  $519^{+18}_{-14}$  &                $2.99\pm0.29$   &  $0.138\pm0.022$\vspace{0.5ex}\\
    MACS~J1720.2+3536    &  0.391  &  $529^{+20}_{-16}$  &                $3.18\pm0.34$   &  $0.132\pm0.016$\vspace{0.5ex}\\
    MACS~J0429.6$-$0253  &  0.399  &  $538^{+40}_{-30}$  &                $3.50\pm0.68$   &  $0.094\pm0.020$\vspace{0.5ex}\\
    MACS~J0159.8$-$0849  &  0.404  &  $621^{+18}_{-20}$  &                $5.54\pm0.59$   &  $0.108\pm0.014$\vspace{0.5ex}\\
    MACS~J2046.0$-$3430  &  0.423  &  $425^{+14}_{-14}$  &                $1.71\pm0.18$   &  $0.166\pm0.023$\vspace{0.5ex}\\
    IRAS~09104+4109      &  0.442  &  $515^{+22}_{-18}$  &                $3.17\pm0.40$   &  $0.096\pm0.012$\vspace{0.5ex}\\
    MACS~J1359.1$-$1929  &  0.447  &  $468^{+25}_{-25}$  &                $2.38\pm0.40$   &  $0.095\pm0.020$\vspace{0.5ex}\\
    RX~J1347.5$-$1145    &  0.451  &  $798^{+30}_{-20}$  &  \hspace{-1ex}$11.79\pm1.14$   &  $0.115\pm0.013$\vspace{0.5ex}\\
    3C~295               &  0.460  &  $447^{+22}_{-16}$  &                $2.12\pm0.28$   &  $0.115\pm0.021$\vspace{0.5ex}\\
    MACS~J1621.3+3810    &  0.461  &  $507^{+16}_{-14}$  &                $3.06\pm0.29$   &  $0.121\pm0.020$\vspace{0.5ex}\\
    MACS~J1427.2+4407    &  0.487  &  $478^{+25}_{-15}$  &                $2.70\pm0.35$   &  $0.144\pm0.017$\vspace{0.5ex}\\
    MACS~J1423.8+2404    &  0.539  &  $472^{+12}_{-10}$  &                $2.67\pm0.19$   &  $0.143\pm0.012$\vspace{0.5ex}\\
    SPT~J2331$-$5051     &  0.576  &  $418^{+30}_{-20}$  &                $2.06\pm0.39$   &  $0.121\pm0.019$\vspace{0.5ex}\\ 
    SPT~J2344$-$4242     &  0.596  &  $592^{+35}_{-35}$  &                $5.75\pm1.01$   &  $0.157\pm0.025$\vspace{0.5ex}\\ 
    SPT~J0000$-$5748     &  0.702  &  $423^{+35}_{-35}$  &                $2.40\pm0.61$   &  $0.081\pm0.023$\vspace{0.5ex}\\ 
    SPT~J2043$-$5035     &  0.723  &  $379^{+10}_{-18}$  &                $1.67\pm0.18$   &  $0.156\pm0.017$\vspace{0.5ex}\\ 
    CL~J1415.2+3612      &  1.028  &  $315^{+12}_{-10}$  &                $1.43\pm0.15$   &  $0.117\pm0.015$\vspace{0.5ex}\\
    3C~186               &  1.063  &  $329^{+22}_{-8}$   &                $1.79\pm0.25$   &  $0.117\pm0.019$\\ 
    \hline
  \end{tabular}
\end{table*}